\documentclass[%
reprint,
 amsmath,amssymb,
 aps,
]{revtex4-2}

\usepackage{graphicx}% Include figure files
\usepackage{dcolumn}% Align table columns on decimal point
\usepackage{bm}% bold math
\usepackage{graphicx}
\usepackage{epstopdf}
\usepackage{algorithmic}
\ifpdf
  \DeclareGraphicsExtensions{.eps,.pdf,.png,.jpg}
\else
  \DeclareGraphicsExtensions{.eps}
\fi
\usepackage[caption=false]{subfig}
\usepackage{amsmath,amssymb,amsfonts}
\usepackage{verbatim}
\usepackage{xcolor}
\usepackage{multirow}
\newcommand{\bs}{\boldsymbol}
\newcommand{\rmd}{\mathrm{d}}
\newcommand{\rmi}{\mathrm{i}}

\begin{document}

\preprint{APS/123-PRA}

\title{Fluorescence profile of a nitrogen-vacancy center in a nanodiamond}% Force line breaks with \\

\author{Qiang Sun}
 \email{qiang.sun@rmit.edu.au}
 \affiliation{%
 Australian Research Council Centre of Excellence for Nanoscale BioPhotonics, School of Science, RMIT University, Melbourne, VIC 3001, Australia
}

\author{Shuo Li}
 \affiliation{%
 Australian Research Council Centre of Excellence for Nanoscale BioPhotonics, School of Science, RMIT University, Melbourne, VIC 3001, Australia
}

%  \altaffiliation[Also at ]{Physics Department, XYZ University.}%Lines break automatically or can be forced with \\
\author{Taras Plakhotnik}%
%  \email{}
\affiliation{%
 School of Mathematics and Physics, The University of  Queensland, QLD 4072, Australia
}%

% \collaboration{MUSO Collaboration}%\noaffiliation

\author{Andrew D. Greentree}
 \email{andrew.greentree@rmit.edu.au}
%  \homepage{http://www.Second.institution.edu/~Charlie.Author}
\affiliation{
 Australian Research Council Centre of Excellence for Nanoscale BioPhotonics, School of Science, RMIT University, Melbourne, VIC 3001, Australia
}%
% \affiliation{
%  Third institution, the second for Charlie Author
% \affiliation{%
%  Authors' institution and/or address\\
%  This line break forced with \textbackslash\textbackslash
% }%

% \collaboration{CLEO Collaboration}%\noaffiliation

\date{\today}% It is always \today, today,

\begin{abstract}
Nanodiamonds containing luminescent point defects are widely explored for applications in quantum bio-sensing such as nanoscale magnetometry, thermometry, and electrometry. A key challenge in the development of such applications is a large variation in fluorescence properties observed between particles, even when obtained from the same batch or nominally identical fabrication processes.  By theoretically modelling the emission of nitrogen-vacancy colour centres in spherical nanoparticles, we are able to show that the fluorescence spectrum varies with the exact position of the emitter within the nanoparticle, with noticeable effects seen when the diamond radius, $a$, is larger than around 100~nm, and significantly modified fluorescence profiles found for larger particles when $a=200$~nm and $a=300$~nm, while negligible effects below $a=100$~nm.  These results show that the reproducible geometry of point defect position within narrowly sized batch of diamond crystals is necessary for controlling the emission properties. Our results are useful for understanding the extent to which nanodiamonds can be optimised for bio-sensing applications.
% \begin{description}
% Secondary publications and information retrieval purposes.
% You may use the \texttt{description} environment to structure your abstract;
% use the optional argument of the \verb+\item+ command to give the category of each item. 
% \end{description}
\end{abstract}

\maketitle

%\tableofcontents

\section{Introduction}

Understanding nanoscale effects is one of the most exciting scientific endeavours. It underpins  very diverse research areas such as the mechanisms of life~\cite{Schechter2008,MYS+2011, Thomas2012, Wang2019}, quantum information~\cite{TARASOV2009, Laucht2021, Heinrich2021}, and fundamental phenomena in condensed matter systems~\cite{Cohen2008, Ou2019, Bachtold2022}.  Research in these areas requires nanoscale quantum sensors, and one of the most mature room-temperature quantum nanoscale sensor is nanodiamond containing the negatively-charged nitrogen-vacancy (NV) centre~\cite{Schirhagl2014, Radtke2019}. %Moreover, nanodiamonds embedded with other point defects such as SiV,GeV  also gain popularity.  
Such doped nanodiamonds are a superb system for quantum sensing. They are highly biocompatible~\cite{Aharonovich2011, Zhu2012} and photostable~\cite{Vaijayanthimala2012, Jung2020}, and therefore are ideal for minimally invasive biological experiments.  

In NV, readout is typically achieved via optically detected magnetic resonance (ODMR), where the resonances in the interaction of the electronic spin of NV centres and a radio frequency (RF) electromagnetic field are detected by measuring the photo luminescence intensity of the centres.  In this way, NV centres have been used for nanoscale magnetometry~\cite{Maze2008, Bai2020}, electrometery~\cite{Dolde2011, Tetienne2017}, thermometry~\cite{Kucsko2013, Khalid2020} and pressure measurements~\cite{Doherty2014}. Alternatively, accurate measurements of the photon luminescence spectrum (in particular its zero-phonon line) allows for all optical measurements~\cite{PDC+2014}.

A drawback of fluorescent nanodiamonds in comparison to quantum dots and organic molecules is their intrinsic heterogeneity. Large variations in fluorescence intensities and lifetimes are observed between NV centres in similar nanodiamonds~\cite{Heffernan2017, Capelli2019, Wilson2019, Capelli2021}.  Understanding the origins of such variations and the ways of reducing the heterogeneity is important for developing a reliable technological platform.  

Here we show large variations in NV fluorescence by performing theoretical modelling of the fluorescence of a point defect in spherical nanodiamonds as a function of nanodiamond size and the defect position within the crystal. To explore the effect of geometry on emission, we treat the NV coupled to  phonons of the crystal lattice  as a set of electric dipoles with different oscillation frequencies and emission probabilities~\cite{Davies1976,Su2008}. The electromagnetic fields within and outside the diamond are calculated using Mie theory~\cite{Mie1908, vandeHulst1957, Bohren1998, Margetis2002} and validated by the numerical solver~\cite{Sun2022}. In our calculations, modification of the density of  states close to crystal surface~\cite{Inam2013} is not considered. Our results show that noticeable variations in the shapes of NV emission spectra are negligible when $a$, the radius of the particle is  below 100~nm but are significantly modified if $a\approx 200$~nm and larger. Although our systems are idealised for computational tractability, the results highlight the sensitivity of fluorescence to the precise location of the NV with diamond crystal, and are therefore important for understanding the experimentally observed variations in fluorescence.

\section{Model}

% \begin{figure}[t]
% \centering{}
% \includegraphics[width=0.45\textwidth]{Fig1_PhysSketch.png}
% \caption{Sketch of the physical model for the photon collections by a pin hole with a circular optical objective emitted from a single NV centre implemented in a spherical nanodiamond.}  \label{fig:sketch}
% \end{figure}

To investigate how the NV centre location within a nanodiamond particle  affects the far field fluorescence, we consider a single NV in a spherical particle with a refractive index of $n_2=2.4$ in air with a refractive index of $n_1=1.0$. The broad NV emission spectrum is represented by emission by 12 point dipoles $\bs{p} \equiv \bs{p}_i \, (i=0,1,2...,11)$ corresponding to the NV emitting a single photon and multiple phonons. This gives rise to a broad emission spectrum with components at different wavelengths, as listed in Table~\ref{tab:brancingratios}. We use the low temperature emission probabilities from Ref.~\cite{Davies1976, Su2008} as the relative intensity, $R$, emitted from the NV centre at different numbers of de-exciting phonons, but we expect similar results for the room temperature case. Since intensity $R$ is proportional to the field power, it is then proportional to the square of the strength of the represented electric dipole for the NV centre. To match the dimension, we have $ (c |\bs{p}|^2) / (4 \pi \epsilon_0 \epsilon_r \lambda^4) \sim R$ where $c$ is the speed of light, $\epsilon_0$ is the vacuum permittivity, $\epsilon_r$ is the relative permittivity and $\lambda$ is the wavelength of emission light. Since $c$, $\epsilon_0$ and $\epsilon_r$ are constant in a homogeneous diamond, for simplicity, we set $|\bs{p}|^2 = \lambda^4 R$. In Fig.~\ref{Fig:10nm} (c-d), the square symbols display the relative intensity $R$ at the corresponding wavelengths.

\begin{figure}[t]
\centering{}
\subfloat[Case A]{ \includegraphics[width=0.24\textwidth]{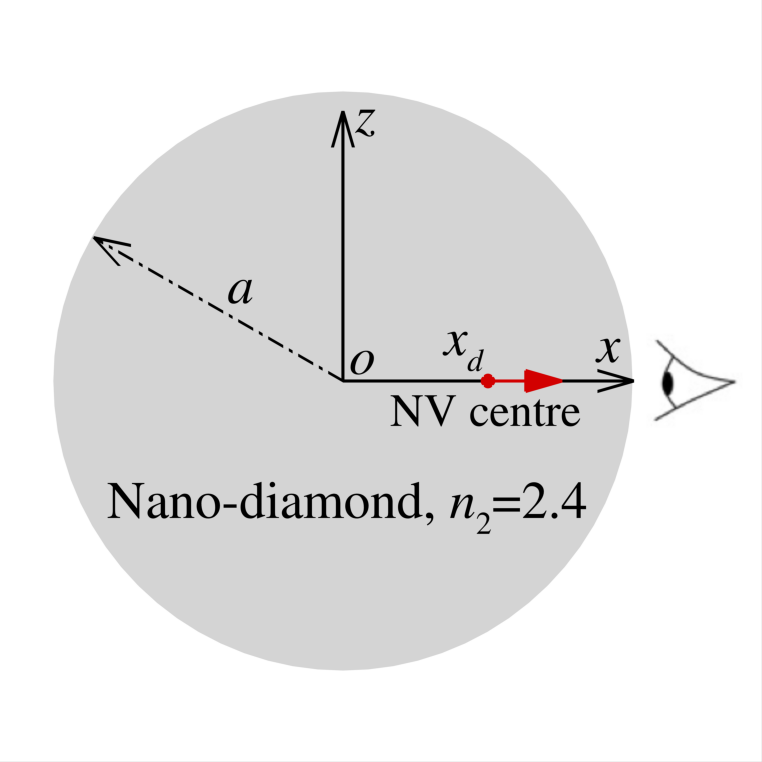}}
\subfloat[Case B]{ \includegraphics[width=0.24\textwidth]{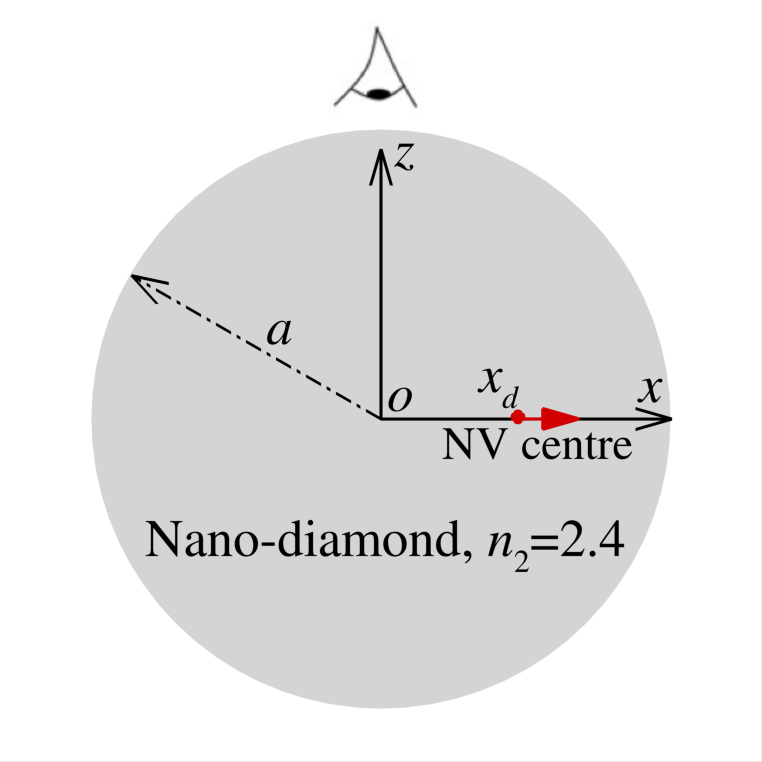}} \\
\subfloat[Case C]{ \includegraphics[width=0.24\textwidth]{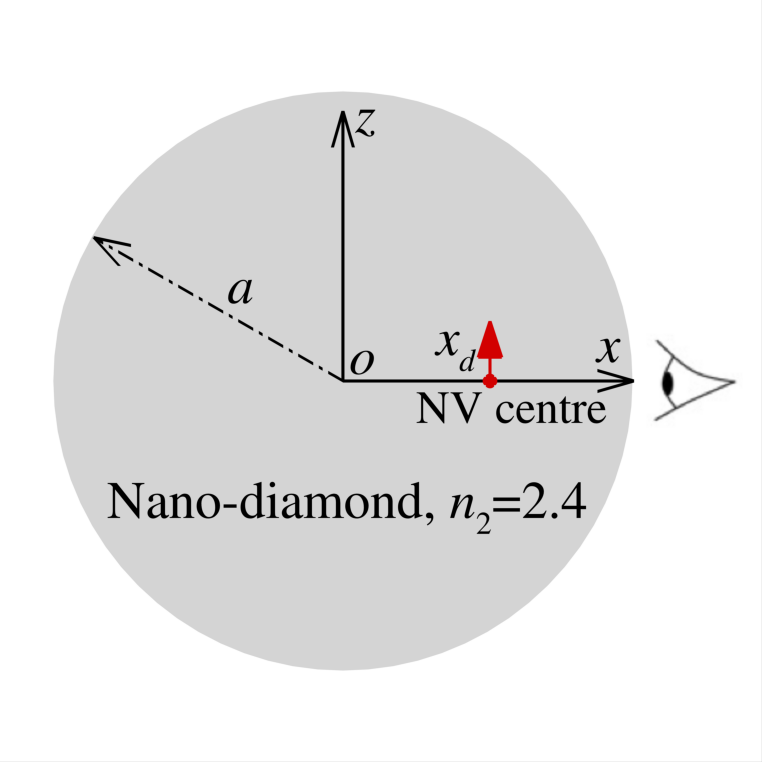}} 
\subfloat[Case D]{ \includegraphics[width=0.24\textwidth]{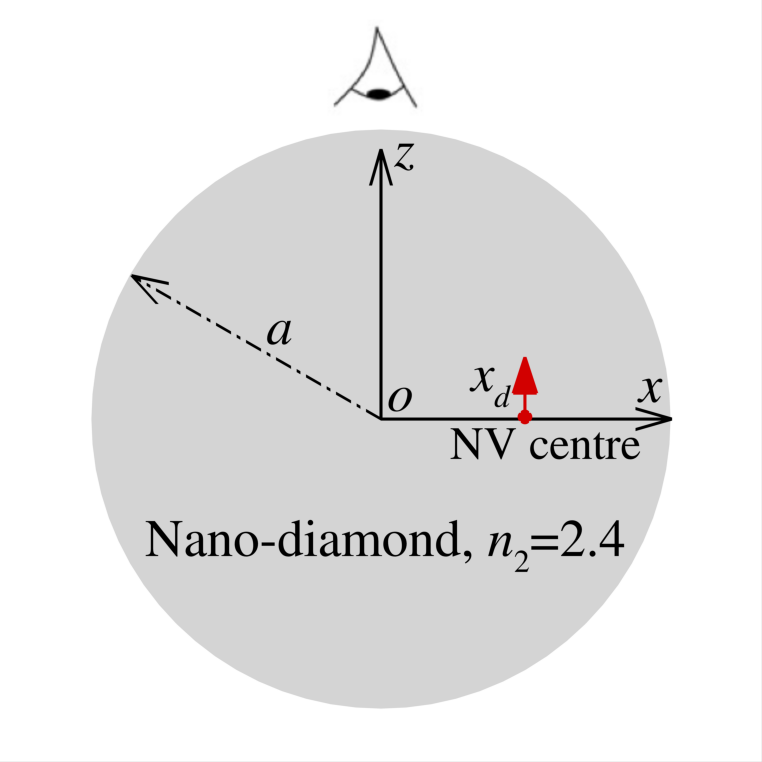}}
\caption{Four cases under consideration for the photon collections by a circular optical objective with NA=0.9 emitted from a single NV centre, which is represented by an electric dipole with moment $\bs{p}$, implemented in a spherical nanodiamond when the NV centre is located at different position along the $x$-axis: (a) Case A, $\bs{p}=(p,0,0)$ and side view; (b) Case B, $\bs{p}=(p,0,0)$ and top view; (c) Case C, $\bs{p}=(0,0,p)$ and side view; and (d) Case D, $\bs{p}=(0,0,p)$ and side view.}  \label{Fig:Cases}
\end{figure}

To monitor the emission, we model a detector with circular entrance aperture (NA=0.9). The axis of the point dipole is assumed either parallel  or perpendicular to the plane of the aperture, as sketched in Fig.~\ref{Fig:Cases}. In a homogeneous medium, the intensity of each wavelength would be proportional to the photon emission probability in the actual spectrum at the same wavelength, which in turn is derived from the emission probabilities. However, the electromagnetic fields transmitted to the surrounding medium (air in this work) are modified due to the boundary conditions on the surface of the particle and can be obtained by solving  Maxwell's equations which are solved using the Mie theory (see Appendices~\ref{sec:appmodel} and~\ref{sec:appsolution}). After obtaining the electromagnetic fields, we can calculate the observed far-field intensity for each dipole as measured through the aperture located either at the top view position  or the side view position. This is done by integrating the time-averaged Poynting vector over the corresponding aperture area:  
\begin{equation}\label{eq:I}
        I_{\text{nd}}(\lambda_{i}) = \int_{S_{\text{obj}}} \frac{1}{2} \left[\bs{E}^{1} \times (\bs{H}^{1})^{*}\right]_\perp \, \rmd S
\end{equation}
with $i=0,1,2...,11$. The above formulations  gives  the  photon count rates relative to the intensities of a NV centre in bulk diamond listed in Table~\ref{tab:brancingratios} in which $\lambda_i$ is the wavelength of the corresponding dipole, superscript asterisk indicates the complex conjugate, and $\perp$ in the subscript shows that the component of the vector product perpendicular to the plane of the aperture. 

We also calculated 
\begin{equation}\label{eq:normalI}
         I_{\text{nd}}^{n}(\lambda_{i})=\frac{I(\lambda_{i})}{\sum^{11}_{i=0} I(\lambda_{i})}
\end{equation}
the normalised spectra which emphasise changes in the shape of the spectra rather than emission strength of the entire spectral band.

\section{Results}

To demonstrate how the position of the NV centre in a spherical nanodiamond can affect the photon collections at the far field, we locate the NV centre at varying positions along the $x$-axis, $\bs{x}_d = (x_d,\,0,\,0)$ and $|x_d| = d$. The equivalent electric dipole moment, $\bs{p}$, can be either along $x$-axis or $z$-axis. Together with two observation spots, the top view and the side view, as shown in Fig.~\ref{Fig:Cases}, we studied four cases: (i) Case A, $\bs{p}=(p,0,0)$ and side view; (ii) Case B, $\bs{p}=(p,0,0)$ and top view; (iii) Case C, $\bs{p}=(0,0,p)$ and side view; and (iv) Case D, $\bs{p}=(0,0,p)$ and side view. Corresponding to Case A to D, the animations of the overall and normalised photon counts for $a=10$~nm to $a=300$~nm when the NV centre is located from the left to the right of the particle are presented in Supp. Mat. 1 to 4 and Supp. Mat. 5 to 8, respectively. Also, the detailed analysis for different size of particles is demonstrated below.

\begin{figure}[t]
\centering{}
\subfloat[Case A]{ \includegraphics[width=0.24\textwidth]{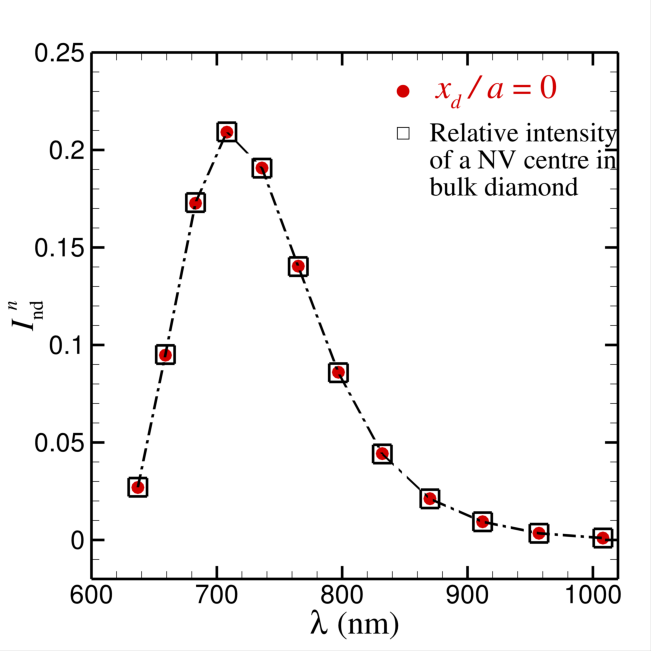}}
\subfloat[Case D]{ \includegraphics[width=0.24\textwidth]{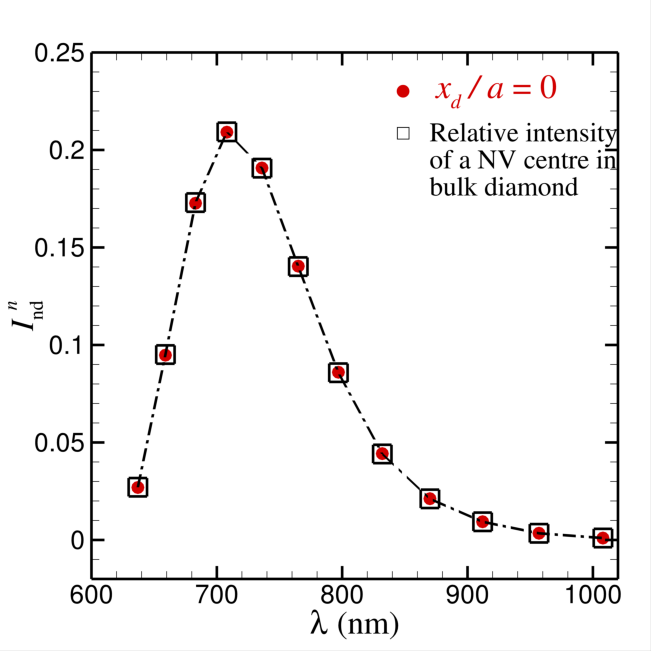}}
\caption{The normalised photon counts (c) and (d) at $x_d/a = 0$ for Case A and Case D, which almost fully represent the relative intensities of a NV centre in bulk diamond at the low-temperature condition listed in Table~\ref{tab:brancingratios}.}  \label{Fig:10nm}
\end{figure}

We start with the case of a small nanodiamond with a radius of $a=10$~nm. When the particle size is small compared to the wavelength of the emitted light from the NV centre, the relative position of the NV centre to the surface of the diamond particle has insignificant effects on the photon collection by the optical objective (the pin hole)~\cite{Plakhotnik2018}, as displayed in Fig.~\ref{Fig:10nm}. One main reason for that is that as the particle size is small, the fields inside the particle is dominated by the near field of the represented electric dipole, and the particle surface is polarised nearly uniformly by such near field profile of the electric dipole. As such, the relative position of the NV centre has negligible effects on photon collection by the optical objective. In Fig.~\ref{Fig:10nm}, we only show the overall and normalised electromagnetic intensity profiles for Case A and Case D as a function of the number of de-exciting phonons, which almost fully represent the relative intensities of a NV centre in bulk diamond at the low-temperature condition listed in Table~\ref{tab:brancingratios}. For Case B and Case C, the profiles are same as what are presented in Fig.~\ref{Fig:10nm}, and hence are not repeated here. 

\begin{figure}[t]
\centering{}
\subfloat[Case A]{ \includegraphics[width=0.24\textwidth]{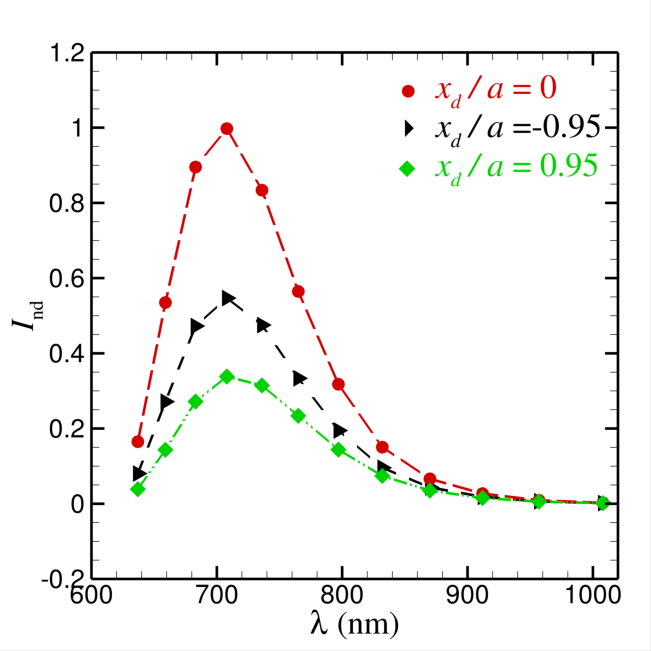}}
\subfloat[Case B]{ \includegraphics[width=0.24\textwidth]{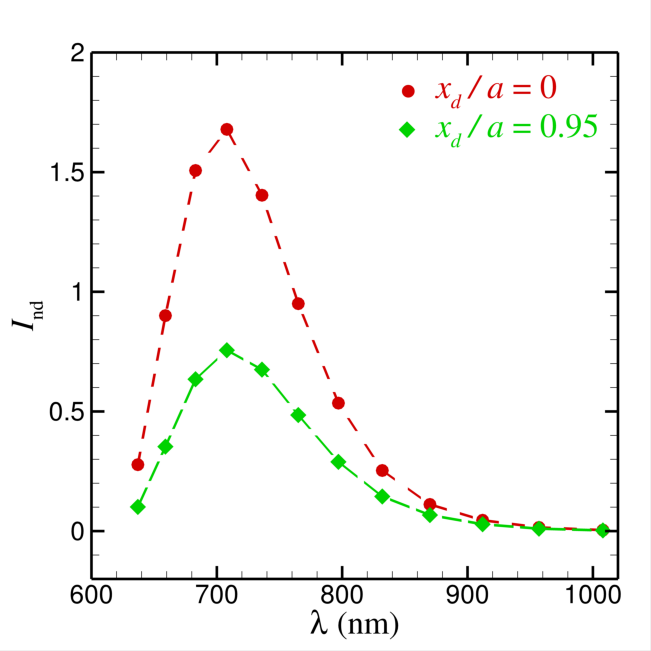}}\\
\subfloat[Case C]{ \includegraphics[width=0.24\textwidth]{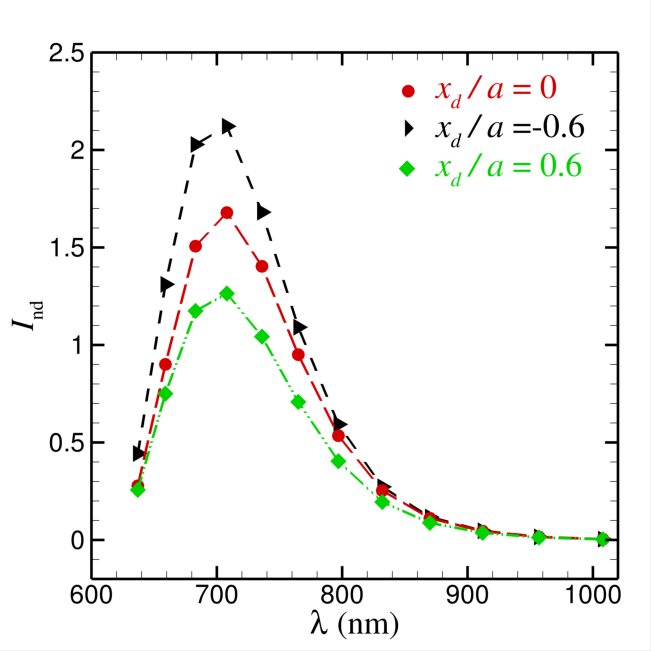}}
\subfloat[Case D]{ \includegraphics[width=0.24\textwidth]{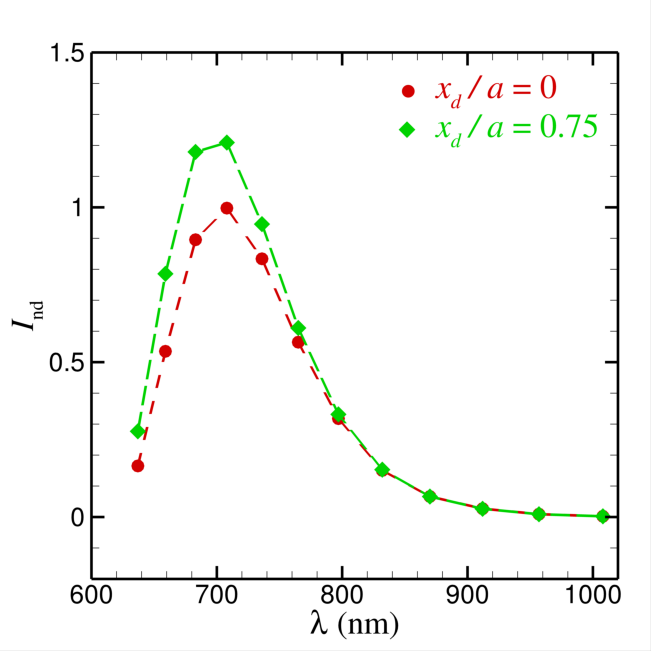}}
\caption{The overall photon counts emitted from a NV centre embedded in a nanodiamond with radius of $a=100$~nm at selected NV centre locations. For the $x$-oriented dipole [(A) side view and (B) top view] the emission is higher when $x_d/a \approx 0$. For the $z$-oriented dipole, the emission is higher when the dipole is close to the diamond particle surface on the left for the side view (Case C) while for the top view (Case D), the emission is weaker when $x_d/a \approx 0$.}  \label{Fig:I100nmSpc}
\end{figure}

When the radius of the diamond particle is 100~nm, the effects on the photon counts emitted from the NV centre due to its location relative to the nanodiamond surface start to present. As the particle size increases, the near field phenomenon from the electric dipole becomes a local effect, and the coupling between the radiation wave from the dipole and particle cavity starts to merge. For example, when the equivalent electric dipole moment direction is along the $x$-axis, the overall electromagnetic field intensity collected by the objective from side (Case A) and top (Case B) view is stronger when the dipole is located in the centre of the diamond particle relative to when it is close to the diamond surface, as shown in Fig.~\ref{Fig:I100nmSpc} (a-b). However, if the dipole moment direction is along the $z$-axis, the overall electromagnetic field intensity is stronger when the dipole is close to the diamond particle surface on the left for the side view, as shown in Fig.~\ref{Fig:I100nmSpc} (c). With the top view for $z$-oriented NV centre, the overall electromagnetic field intensity profile is symmetric with respect to the centre of the diamond centre. The emission is weaker when the dipole is near the centre of the particle relative to  when it is close to the particle surface, as displayed in Fig.~\ref{Fig:I100nmSpc} (d).

\begin{figure}[t]
\centering{}
\subfloat[Case A]{ \includegraphics[width=0.24\textwidth]{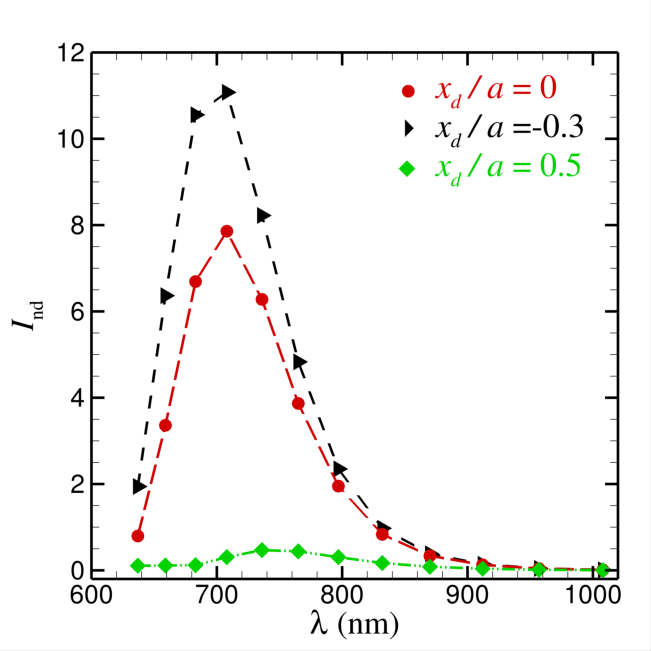}}  
\subfloat[Case B]{ \includegraphics[width=0.24\textwidth]{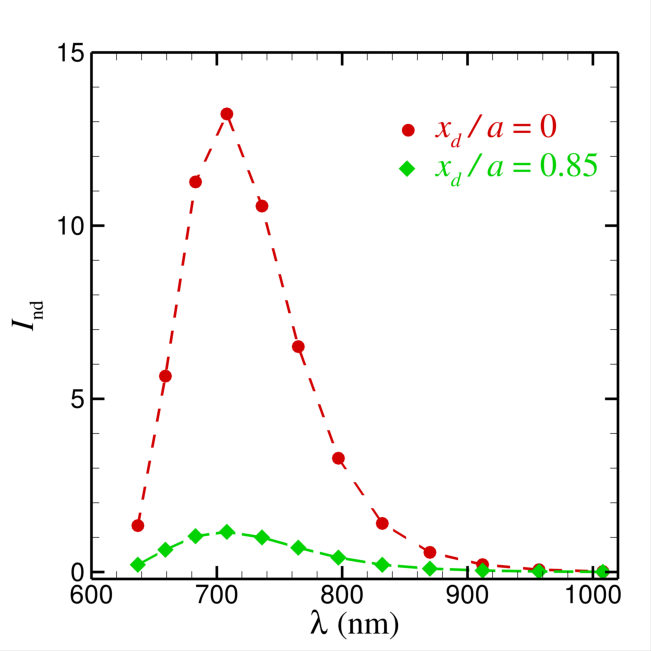}} \\
\subfloat[Case C]{ \includegraphics[width=0.24\textwidth]{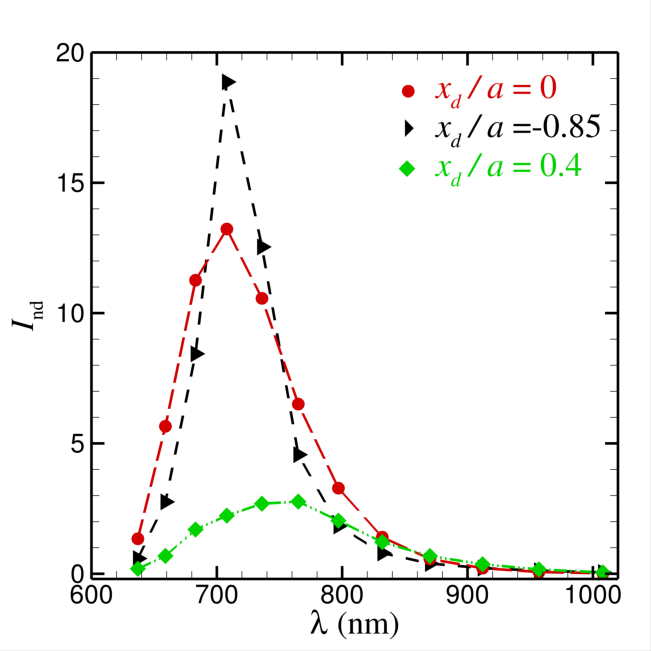}}  
\subfloat[Case D]{ \includegraphics[width=0.24\textwidth]{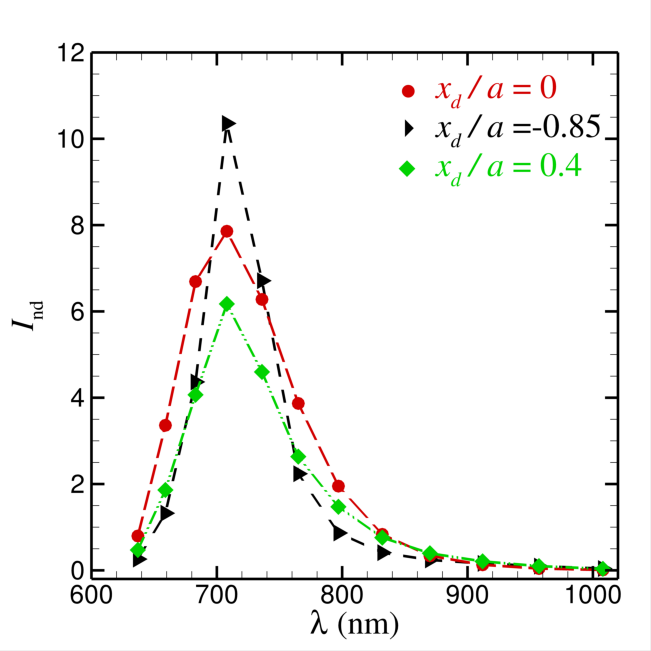}}
\caption{The overall photon counts emitted from a NV centre embedded in a nanodiamond with radius of $a=200$~nm at selected NV colour centre locations. From the side view [Case A and C], the emission is stronger as $x_d/a < 0$ with both $x$- and $z$-oriented dipoles. From the top view, the emission is much enhanced when $x_d/a \approx 0$ for the $x$-oriented dipole [Case B].}  \label{Fig:I200nmSpc}
\end{figure}

When the radius of the diamond particle is 200~nm, the subtle effects that were predicted for the 100~nm particles become far more pronounced. Large changes in both the overall and relative (normalised) spectra are observed. The spectra for the overall electromagnetic field intensity for the four cases are shown in Fig.~\ref{Fig:I200nmSpc}. If the equivalent electric dipole moment direction is along the $x$-axis, the overall electromagnetic field intensity collected from both the top and side views indicate that, when the NV centre is deep in the nanodiamond particle, the fluorescence signals are much stronger than that when it is close to the particle surface, as shown in Fig.~\ref{Fig:I200nmSpc} (a-b). Unlike the symmetric fluorescence profile from the top view, the nanodiamond is much brighter when the NV centre locates in the left part of the particle ($x_d/a < 0$) from the comparison between $x_d/a=-0.3$ and $x_d/a=0.5$ in Fig.~\ref{Fig:I200nmSpc} (a). Whereas if the dipole moment direction is in $z$-direction, for example, Case C and D in Fig.~\ref{Fig:I200nmSpc} (c-d), emission signals from the NV centre is significant when it is either close to the particle surface or near the centre of the diamond particle. 

\begin{figure}[t]
\centering{}
\subfloat[Case A]{ \includegraphics[width=0.24\textwidth]{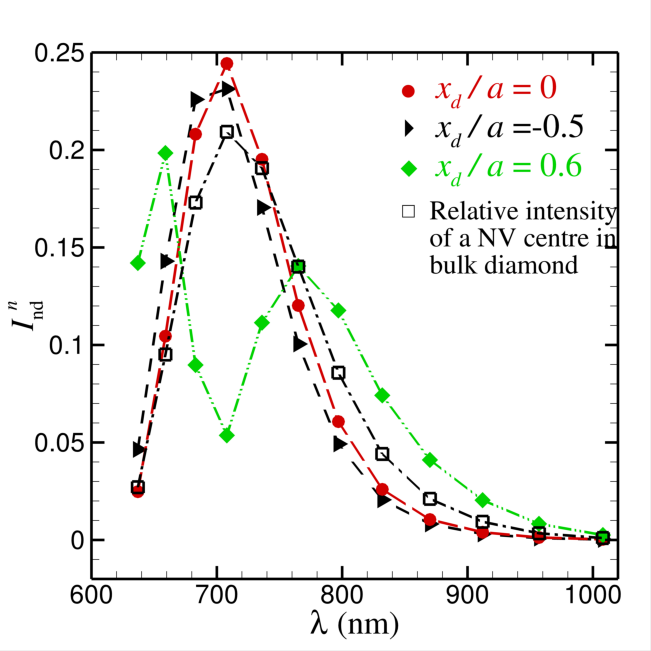}}  
\subfloat[Case B]{ \includegraphics[width=0.24\textwidth]{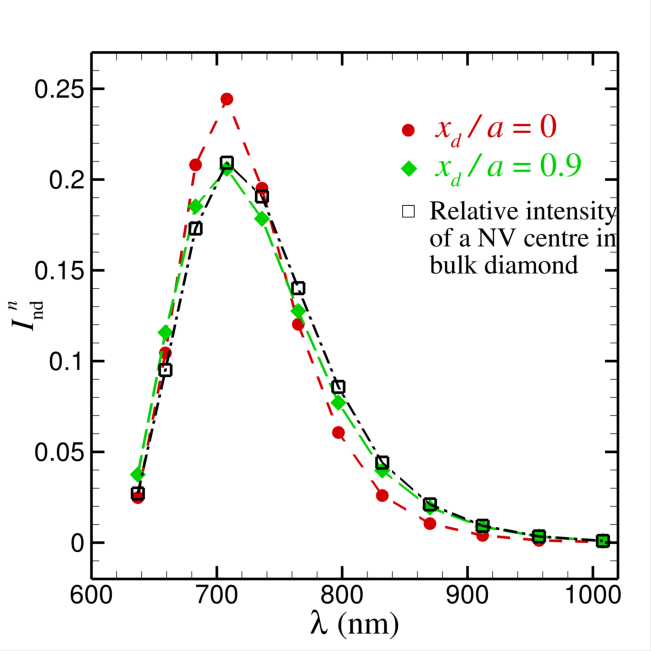}} \\
\subfloat[Case C]{ \includegraphics[width=0.24\textwidth]{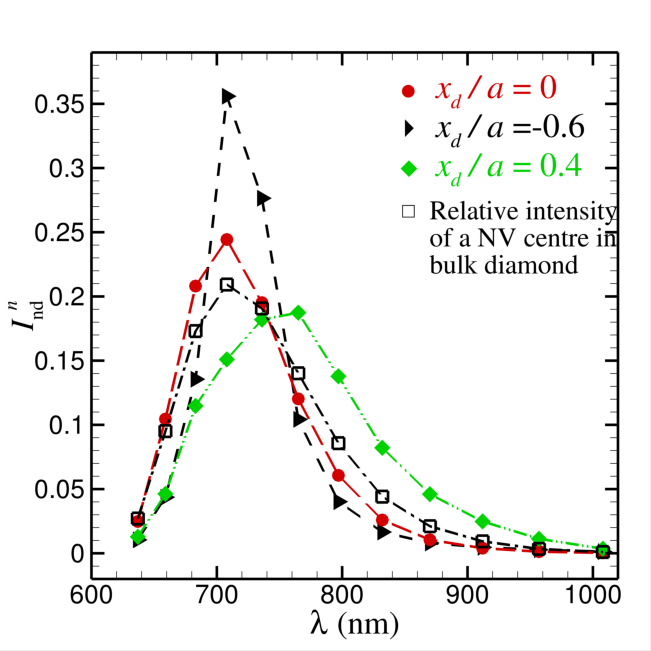}}  
\subfloat[Case D]{ \includegraphics[width=0.24\textwidth]{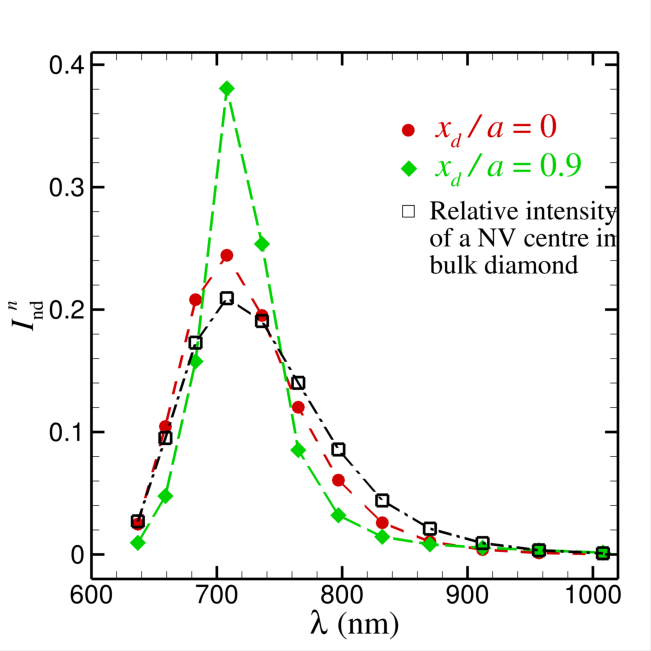}}
\caption{The normalised photon counts emitted from a NV centre embedded in a nanodiamond with radius of $a=200$~nm at selected NV colour centre location. For the $x$-oriented dipole from the side view [Case A], when $x_d/a=0.6$, the original emission peak at 709 nm disappears while two peaks appear at 659 nm and 765 nm.}  \label{Fig:In200nmSpc}
\end{figure}

The normalised electromagnetic field intensity profiles of a single NV centre implemented in a nanodiamond with radius of 200~nm are shown in Fig.~\ref{Fig:In200nmSpc}. For Case A when the dipole moment is along $x$-axis and the photon collection is along the side view, the normalised electromagnetic field intensity almost represents the relative intensities of a NV centre in bulk diamond when the NV centre is located in the left part of the nanodiamond particle ($x_d/a<0$). Nevertheless, if the NV centre is placed to the right part in the nanodiamond when $x_d/a>0$, compared to the relative intensities of a NV centre in bulk diamond, dominant wavelength of the normalised electromagnetic field intensity collected from the side view is firstly changes from $\lambda=$~708~nm to $\lambda=$~736~nm at around $x_d/a = 0.5$ and then changes again to $\lambda=$~659~nm at around $x_d/a = 0.6$, as shown in Fig.~\ref{Fig:In200nmSpc} (a). Also, at $x_d/a = 0.6$, there is a second peak of the normalised electromagnetic field intensity at $\lambda=$~765~nm, while the signal at 708~nm is significantly reduced. Regarding to the top view as presented in Fig.~\ref{Fig:In200nmSpc} (b) for Case B, the normalised electromagnetic field intensity profile is similar to that of a NV centre in bulk diamond when the NV centre is located from side to side in the particle. If the dipole moment direction is $z$-oriented, both the side and top views show that the emission signal is enhanced significantly when the NV centre is close to the surface of the particle ($|x_d|/a > 0.5$) for the wavelength at $\lambda=$~708~nm, as shown in Fig.~\ref{Fig:In200nmSpc} (c-d). When the $z$-oriented NV centre is deep in the particle, from the side view, the dominant number of de-exciting phonons changes from three ($\lambda=$~708~nm) to five ($\lambda=$~765~nm) around $x_d/a = 0.4$, as shown in Fig.~\ref{Fig:In200nmSpc} (c).

\begin{figure}[t]
\centering{}
\subfloat[Case A]{ \includegraphics[width=0.24\textwidth]{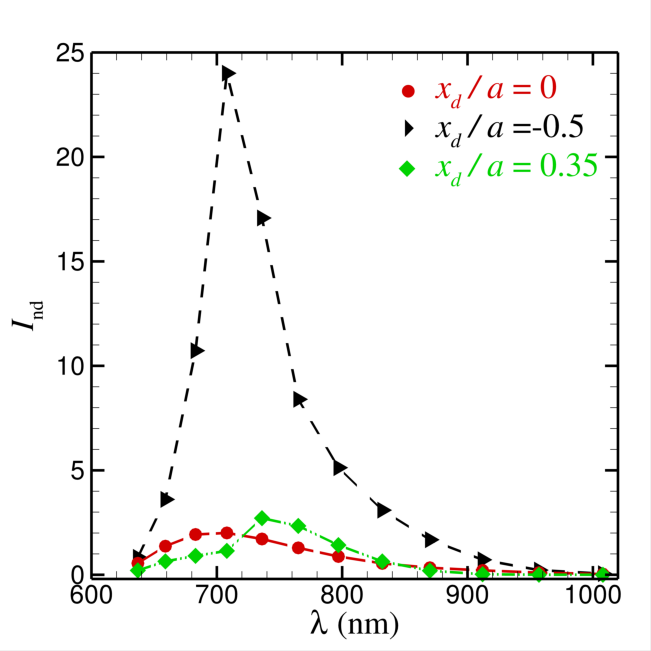}}  
\subfloat[Case B]{ \includegraphics[width=0.24\textwidth]{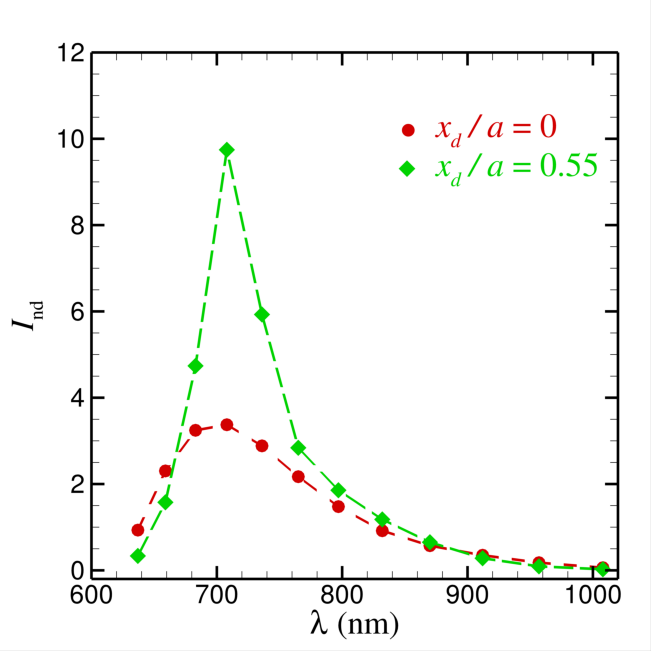}} \\
\subfloat[Case C]{ \includegraphics[width=0.24\textwidth]{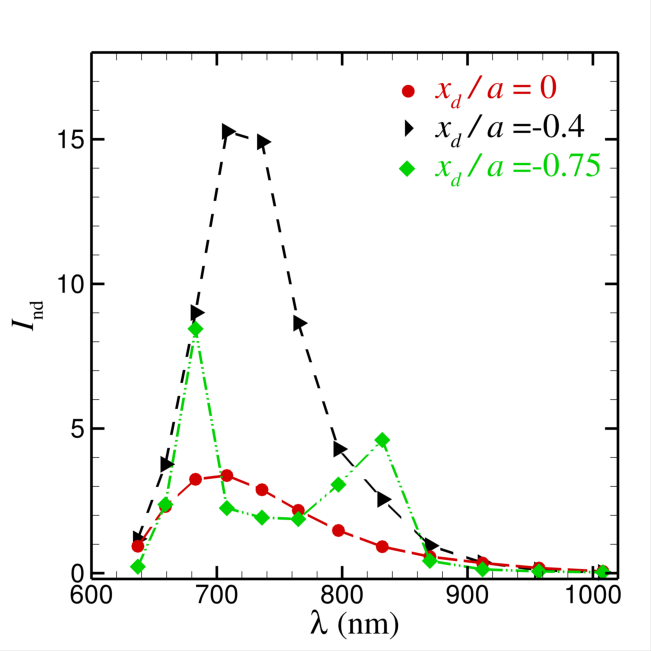}}  
\subfloat[Case D]{ \includegraphics[width=0.24\textwidth]{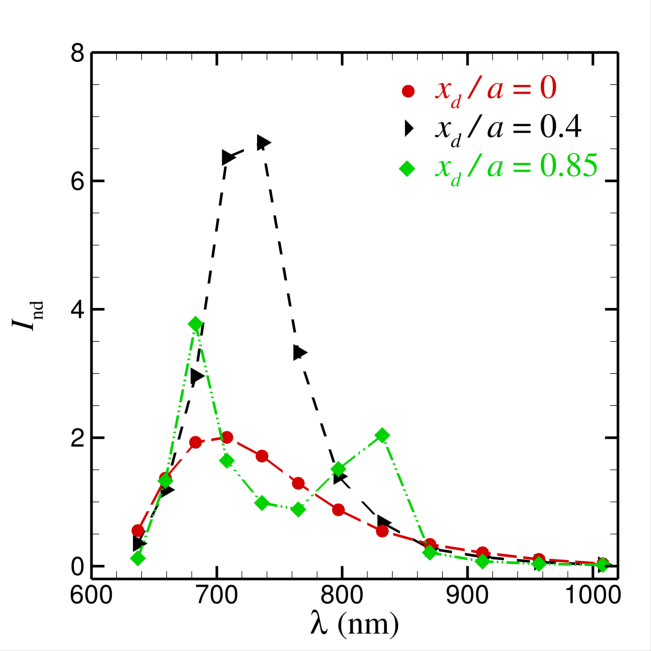}}
\caption{The overall photon counts emitted from a NV centre embedded in a nanodiamond with radius of $a=300$~nm at selected NV colour centre locations. For the $x$-oriented dipole, the emission is strongly enhanced when $x_d/a=-0.5$ from the side view [Case A] and when $x_d/a = 0.55$ from the top view [Case B]. For the $z$-oriented dipole, the emission is highly enhanced when $x_d/a=-0.4$ from the side view [Case C] and when $x_d/a = 0.4$ from the top view [Case D]. Also, the emission spectrum profiles are changed significantly when $x_d/a=-0.75$ from the side view and when $x_d/a= 0.85$ from the top view.}  \label{Fig:I300nmSpc}
\end{figure}

As the diamond radius increases to 300~nm, the spectra become richer. This is because there are numerous opportunities for resonances over the various wavelengths. For a diamond particle with radius of 300~nm, if the equivalent electric dipole moment direction is in the $x$-direction, the overall electromagnetic field intensity at $\lambda=$~708~nm is much stronger when the NV centre is around $x_d/a = -0.5$ in the particle from the side view, as shown in Fig~\ref{Fig:I300nmSpc} (a) for Case A. From the top view, the symmetric profile of the field intensity with respect to the particle centre is obtained when the NV centre is located from one side to the other of the particle, and the strongest fluorescence signal happens at $|x_d|/a = 0.55$ for $\lambda = 708$~nm, as shown in Fig~\ref{Fig:I300nmSpc} (b) for Case B. When the dipole moment direction is pointing along the $z$-axis, the highest fluorescence signal happens at $x_d/a = -0.4$ for $\lambda=$~708~nm from the side view, as shown in Fig~\ref{Fig:I300nmSpc} (c), while from the top view, the electromagnetic field intensity profile is symmetric to the particle centre and the strongest appears at around $|x_d|/a = 0.4$ for $\lambda=$~708~nm and $\lambda=$~736~nm. Also, for these two cases, when $x_d/a=-0.75$ from the side view and $|x_d|/a = 0.85$ from the top view, there are two peaks of the fluorescence signals at $\lambda=683$~nm and $\lambda=832$~nm while the original peak signal at $\lambda=708$~nm for a NV centre in bulk diamond is significantly reduced. 

\begin{figure}[t]
\centering{}
\subfloat[Case A]{ \includegraphics[width=0.24\textwidth]{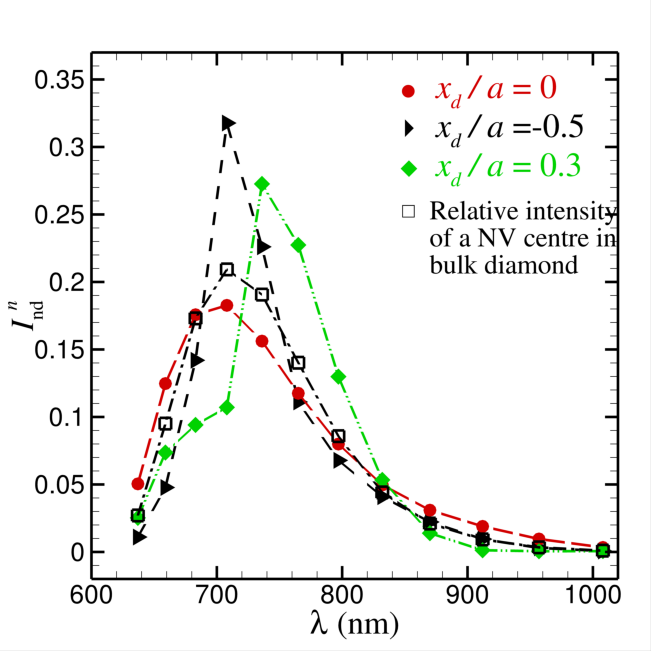}}  
\subfloat[Case B]{ \includegraphics[width=0.24\textwidth]{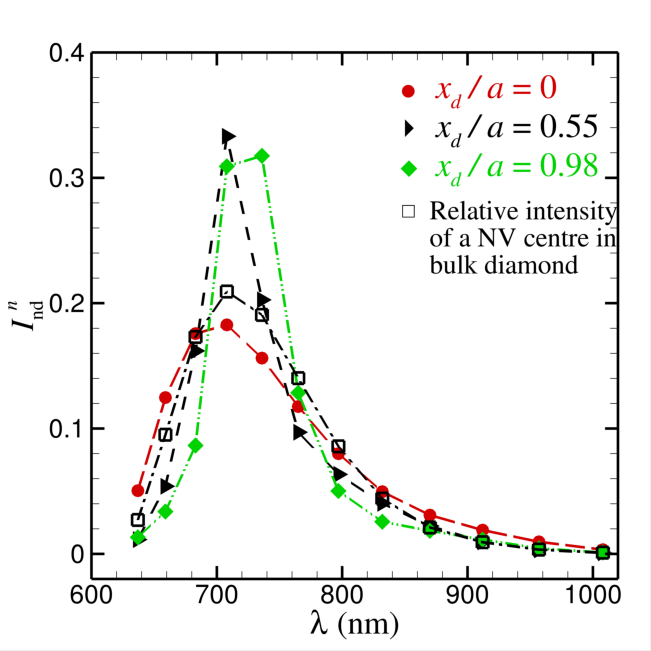}} \\
\subfloat[Case C]{ \includegraphics[width=0.24\textwidth]{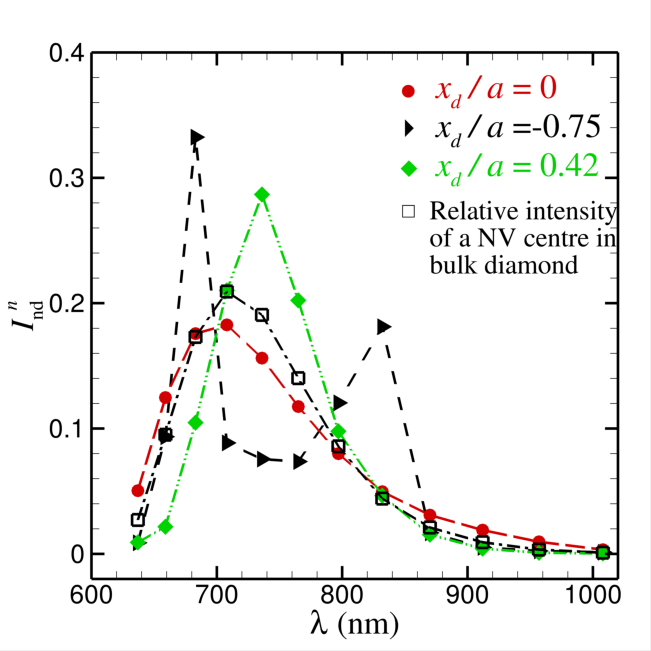}}  
\subfloat[Case D]{ \includegraphics[width=0.24\textwidth]{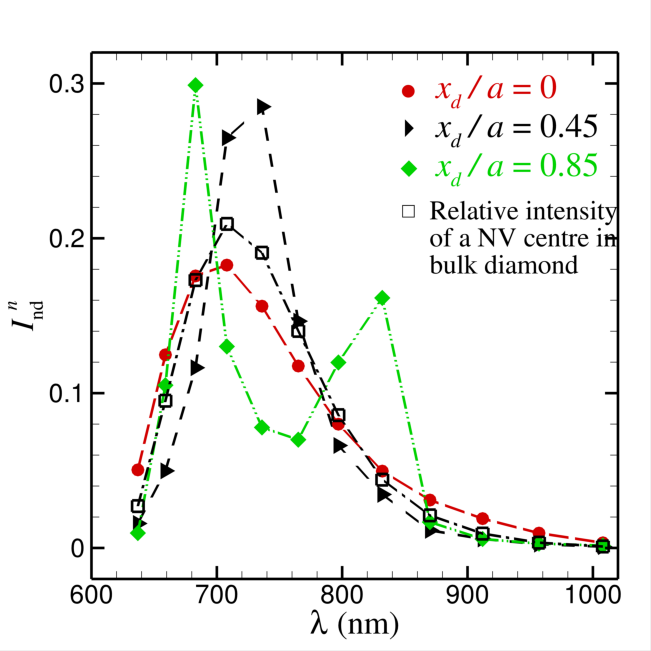}}
\caption{The normalised photon counts emitted from a NV centre embedded in a nanodiamond with radius of $a=300$~nm at selected NV colour centre locations. For the $z$-oriented dipole, the emission spectrum profiles are changed significantly relative to that of a NV centre in a bulk diamond, for example when $x_d/a=-0.75$ from the side view [Case C] and when $x_d/a= 0.85$ from the top view [Case D].}  \label{Fig:In300nmSpc}
\end{figure}

On the normalised fluorescence signals when a NV centre is located at different position in a diamond particle with radius of 300~nm, for Case A and B when the electric dipole moment direction is along $x$-axis, the dominant emission fluorescence is the same as a NV centre in bulk diamond at $\lambda=$~708~nm when the NV centre locates close to the surface of the diamond particle, as shown in Fig~\ref{Fig:In300nmSpc} (a-b). When the NV centre locates at $x_d/a=0.3$, the dominant emission wavelength changes to $\lambda=$~736~nm, as shown in Fig.~\ref{Fig:In300nmSpc} (a). For Case C and D as the dipole moment direction is in $z$-direction, when the position of the NV centre is close to the surface of the diamond particle, the strongest emission happens at $\lambda=$~683~nm relative to a NV centre in bulk diamond at $\lambda=$~708~nm, as shown in Fig~\ref{Fig:In300nmSpc} (c) and (d). If the NV centre location locates deeper in the diamond particle at around $x_d/a=0.45$, the dominant emission is changed to $\lambda=$~736~nm. Also, when $x_d/a=-0.75$ from the side view and at $|x_d|/a = 0.8$ from the top view for a $z$-oriented NV centre, there are the two peaks of the normalised fluorescence signals at $\lambda=683$~nm and $\lambda=832$~nm while the original peak signal at $\lambda=708$~nm for a NV centre in bulk diamond is significantly reduced. 

When comparing the fluorescence profiles from a 300~nm diamond to those from the smaller diamonds, the emission from longer wavelengths are enhanced in the 300~nm case. This is because the particle size at radius of 300~nm is comparable to the longer wavelengths when the high refractive index of diamond is taken into consideration, which leads to the enhanced cavity effects of the diamond particle for the emission at the higher order lines~\cite{Almokhtar2014}. 

\section{Discussion}

There are basically two effects on the emission spectra. One of them is the change of the integrated intensity, represented in Eq.~(\ref{eq:I}), and the other is the change of the normalized spectra, represented in Eq.~(\ref{eq:normalI}). Both effects depend on the position of the NV centre within the crystal defined by $x_d$, orientation of its transition dipole moment, and on the crystal particle size $a$. Qualitatively, the variation of the overall intensity is negligible if $2\pi a \lambda \ll 1$ and the variation of the normalized spectra is negligible if $2\pi a \delta \lambda / \lambda^2 \ll 1$ where $\delta \lambda $ is the width of the luminescence spectrum of the centre. For NV-centres, the value $\delta \lambda / \lambda \sim 1/7 $ and therefore the change in the normalised spectra is observed for significantly larger crystals.

\begin{figure}[t]
\centering{}
\subfloat[]{ \includegraphics[width=0.24\textwidth]{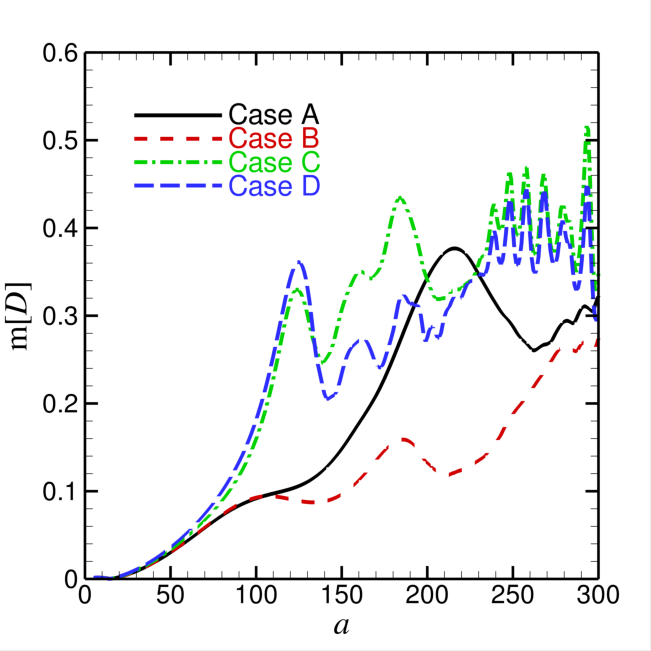}}  
\subfloat[]{ \includegraphics[width=0.24\textwidth]{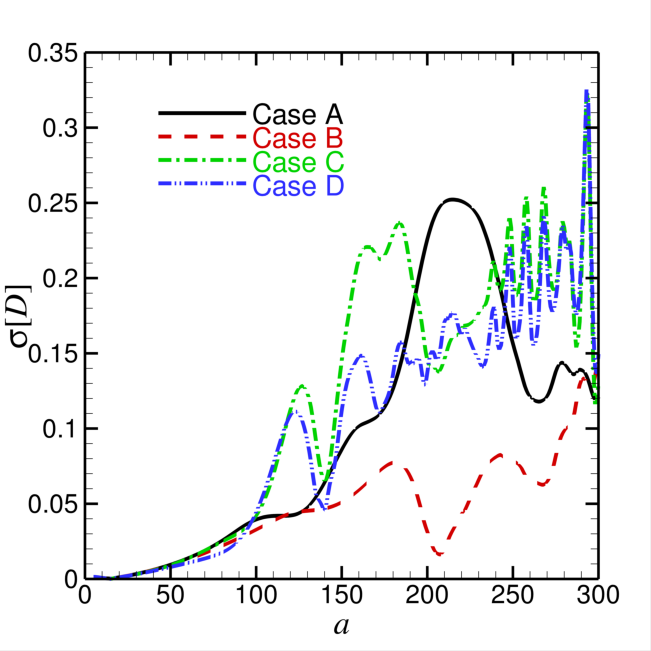}}
\caption{Comparisons of the normalised emission spectra of a NV centre in diamond particles to that in a bulk diamond.}  \label{Fig:meanDstdD}
\end{figure}

It would be worth monitoring how the normalised emission spectra of a NV center in a diamond particle differ from that in a bulk diamond crystal. To characterise the difference, we calculate a value $D$ defined
as follows
\begin{equation}
    D(x_d,a) \equiv P_{D}(x_d,a) \sum_{i=0}^{11} \left|I^n_{\text{nd}}(\lambda_{i},x_d) - I^n_{\text{bulk}}(\lambda_{i})\right|,
\end{equation}
in which $I^n_{\text{nd}}(\lambda_{i},x_d)$ is defined in Eq.~(\ref{eq:normalI}), and the values of $I^n_{\text{bulk}}(\lambda_{i})$ are the relative intensity $R$ listed in \ref{tab:brancingratios}. When computing $D$, we also take into account of the possibility to implement a NV centre with respect to the location $x_d$, which is denoted as $P_{D}$. If location $x_d$ is close to the surface, such as $a-|x_d| \le 2$ nm, it is nearly impossible to implement NV centres. While if $x_d$ is deep inside the particle, such as $a-|x_d| \ge 5$ nm, the chance to implement a NV centre is fairly the same. As such, the value $P_{D}$ can be represented by an error function depending $a-|x_d|$ with mean value of $3$ nm and standard derivation of $0.5$:
\begin{equation}
    P_{D}(x_d,a) \equiv \frac{1}{2}\left[1+\text{erf}\left(\frac{2(a-|x_d|-3)}{\sqrt{2}}\right)\right].
\end{equation}
For each particle with radius of $a$, we compute the mean value and the stand derivation of $D$ with respect to $x_d$ to characterise how different the normalised emission spectra is from that of a bulk diamond. As shown in Fig.~\ref{Fig:meanDstdD}, along with the increase of the particle size, the overall trend of the mean value of $D$ grows, which indicates that the normalised emission spectra of a NV centre are more likely different in larger particles than smaller ones relative to the emission spectrum of a NV centre in bulk. Also, for the $z$-oriented dipole (Case C and D), the mean value of $D$ oscillates significantly along with particle size when $a>230$ nm. The standard derivation of $D$ has a similar trend as its mean value except for the particle size at around $a=210$ nm with the $x$-oriented dipole (Case A and B). For Case A when the particle size is $a=210$ nm, the $x$-oriented dipole with the side view, the difference between the normalised emission spectra of a NV centre in a particle relative to that in bulk is significantly depends on the location of the dipole $x_d$. However, from the top view (Case B), the effects of the location of the dipole is negligible.

\section{Conclusion}
We performed theoretical modelling of the fluorescence profiles of a NV colour centre in a spherical nanodiamond, exploring the effects of the relative location, orientation, and nanodiamond size on the emission probabilities of NV centre together. Changes in the emission probabilities lead to variations in the expected fluorescence profile. Our calculations indicate that the effects of the relative location, orientation of NV centre on the fluorescence signals become noticeable when the particle radius is greater than around $a=100$~nm and much profound for larger particles when $a=200$ nm and $a=300$ nm, with negligible effects below $a=100$~nm. Our results indicate that the information of the exact geometry of NV-diamond system is critical to understand and control the fluorescence profile, which is of importance to optimise such systems for quantum bio-sensing applications.

\begin{acknowledgments}
Q.S. acknowledges the support from the Australian Research Council grant DE150100169. A.D.G. acknowledges the support from the Australian Research Council grant FT160100357. Q.S., S.L. and A.D.G. acknowledge the Australian Research Council grant CE140100003. S.L. and A.D.G acknowledge the Air Force Office of Scientific Research (FA9550-20-1-0276). This research was partially undertaken with the assistance of resources from the National Computational Infrastructure (NCI Australia), an NCRIS enabled capability supported by the Australian Government (Grant No. LE160100051).
\end{acknowledgments}

\appendix
\renewcommand\thefigure{\thesection.\arabic{figure}} 
\renewcommand\thetable{\thesection.\arabic{table}}

\section{Theoretical model}\label{sec:appmodel}
\setcounter{figure}{0}  
\setcounter{table}{0}  

\begin{table*}[t]
\centering
\caption{Probabilities (Relative intensity $R$) to emit photons and phonons of a NV centre in bulk diamond as a function of the number of phonons at low temperature calculated in Refs.~\cite{Davies1976}. The zero phonon line is indicated by ZPL, and the phononic sideband arises from the summation from 1 to 11 phonons.}
%\resizebox{1.0\textwidth}{!}{
\begin{tabular}{| c | c | c | c |}
\hline
No. of phonons & Wavelength $\lambda$  & Emission probabilities  & Dipole moment strength (arb. u.) \\
 & (nm) & (Relative intensity $R$) & ($|p| = \lambda^2 \sqrt{R}$)  \\ 
\hline
0  (ZPL) & 637 & 0.0270 & 66674.65 \\
1  & 659  & 0.0951  & 133924.83 \\
2  & 683  & 0.173  & 194028.02 \\
3 & 708  & 0.209   & 229160.45 \\
4  & 736  & 0.191  & 236740.36 \\
5  & 765  & 0.140  & 218971.14 \\
6  & 797  & 0.0856 &  185846.13 \\
7  & 832  & 0.0441 & 145367.04 \\
8  & 870  & 0.0211 &  109946.08 \\
9  & 912  & 0.00931 & 80253.60 \\
10 & 957  & 0.00343 & 53637.80 \\
11  & 1008 & 0.000980  & 31807.83 \\
\hline
\end{tabular}
%}
\label{tab:brancingratios}
\end{table*}

In our model, we represent a single NV in a spherical particle with a refractive index of $n_2=2.4$ by an electric dipole. The broad NV emission spectrum is represented by emission by 12 point dipoles $\bs{p} \equiv \bs{p}_i \, (i=0,1,2...,11)$ corresponding to the NV de-exciting via a single photon and multiple phonons as listed in Table~\ref{tab:brancingratios}. All electric dipoles are co-located at $\bs{x}_d$ but each of them oscillates at a specific angular frequency $\omega \equiv \omega_i \, (i=0,1,2...,11)$ as $\exp{(-\rmi \omega t)}$. The emitted electric and magnetic fields from such a dipole are, respectively,
\begin{subequations}\label{eq:EMfielddipole}
    \begin{align}
        \bs{E}^{d} &= \frac{1}{4\pi \epsilon_0 \epsilon_2}\frac{\exp( \rmi k_2 r_d)}{ r_d^3}\Bigg\{(-k_2^2 r_d^2 - 3 \rmi k_2 r_d +3) \frac{\boldsymbol r_d \cdot \boldsymbol p}{r_d^2}\boldsymbol r_d  \nonumber \\
        &\qquad \qquad \qquad \qquad  + (k_2^2 r_d^2 + \rmi k_2 r_d -1) \boldsymbol p \Bigg\}, \\
        \bs{H}^{d} &= \frac{\omega k_2}{4\pi}[\bs{r}_d \times \bs{p}] \left( \frac{1}{r_d} - \frac{1}{\rmi k_2 r_d^2} \right) \frac{\exp( \rmi k_2 r_d)}{r_d}
    \end{align}
\end{subequations}
where $\bs{r}_d = \bs{x} - \bs{x}_d$ with $\bs{x}$ being the field location of interest and $r_d = |\bs{r}_d|$, $k_2$ is the wavenumber, $\epsilon_0 $ is the permittivity in vacuum, and $\epsilon_2 = n_2^2$ is the relative permittivity of diamond with $n_2 = 2.4$ the refractive index of diamond. 

In a homogeneous medium, the intensity of each wavelength would be proportional to the photon emission probability in the actual spectrum at the same wavelength, which in turn is derived from the emission probabilities. However, the electromagnetic fields transmitted to the surrounding medium (air in this work) are modified due to the boundary conditions on the surface of the particle and can be obtained by solving  Maxwell's equations. In the frequency domain, the Maxwell's equations in the internal domain of the nanodiamond and the external domain are
\begin{subequations}\label{eq:Maxwelleq}
    \begin{align}
        \nabla \times \boldsymbol E^{j} &= \rmi \omega \mu_0 \mu_j \bs{H}^{j}, \\
        \nabla \cdot \boldsymbol E^{j} & = 0; \\
        \nabla \times \boldsymbol H^{j} &=-\rmi \omega \epsilon_0 \epsilon_j \bs{E}^{j}, \\
        \nabla \cdot \boldsymbol H^{j} &= 0
    \end{align}
\end{subequations}
where $\mu_0$ is the permeability in vacuum, $j$ refers to the external domain and the nanodiamond domain with $j=1$ and $j=2$, respectively, and $\mu_j$ is the relative permeability of each domain which is set as $\mu_1 = \mu_2 = 1$ in this work. 

Together with the boundary conditions,
\begin{subequations}
    \begin{align}
        &\bs{t}_1 \cdot (\bs{E}^{2} + \bs{E}^{d}) = \bs{t}_1\cdot \bs{E}^{1}, \quad \bs{t}_2 \cdot (\bs{E}^{2} + \bs{E}^{d}) = \bs{t}_2\cdot \bs{E}^{1}; \\
        &\bs{t}_1 \cdot (\bs{H}^{2} + \bs{H}^{d}) = \bs{t}_1\cdot \bs{H}^{1}, \quad \bs{t}_2 \cdot (\bs{H}^{2} + \bs{H}^{d}) = \bs{t}_2\cdot \bs{H}^{1}
    \end{align}
\end{subequations}
where $\bs{t}_1$ and $\bs{t}_2$ are the two independent unit tangential directions on the diamond surface, the Maxwell's equations~(\ref{eq:Maxwelleq}) are solved using the Mie theory, which is detailed in the Appendix~\ref{sec:appsolution}. 

\section{Solution for the electromagnetic fields emitted from a NV centre in a spherical diamond particle}~\label{sec:appsolution}
\setcounter{figure}{0}  
\setcounter{table}{0} 

The solution procedure to calculate the electromagnetic fields emitted from a NV centre in a spherical diamond particle is given. It is worth noting that to easily and clearly show the calculation procedure and apply the usual setup of a spherical coordinate system, the equivalent electric dipole for the NV centre is chosen to locate along the the axis of symmetry ($z$ axis) from which the polar angle is measured in this appendix. It is straightforward to use the solution given here to get the results presented in the main text via simple coordinate transform and rotation. 

\begin{figure}[b]
\centering{}
\subfloat[Vertical electric dipole]{ \includegraphics[width=0.24\textwidth]{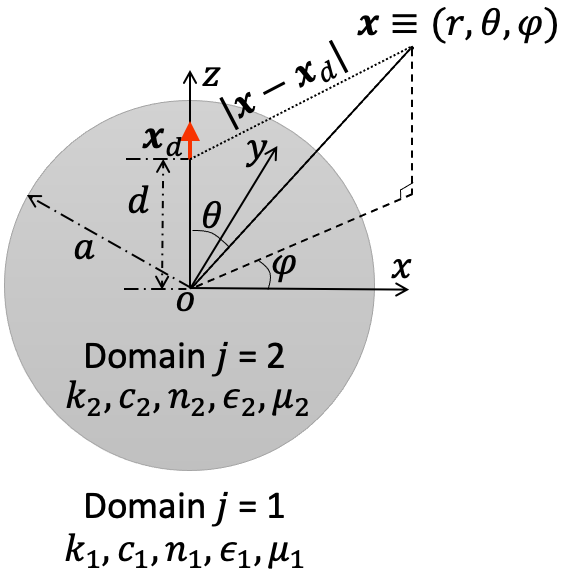}}  
\subfloat[Horizontal electric dipole]{ \includegraphics[width=0.24\textwidth]{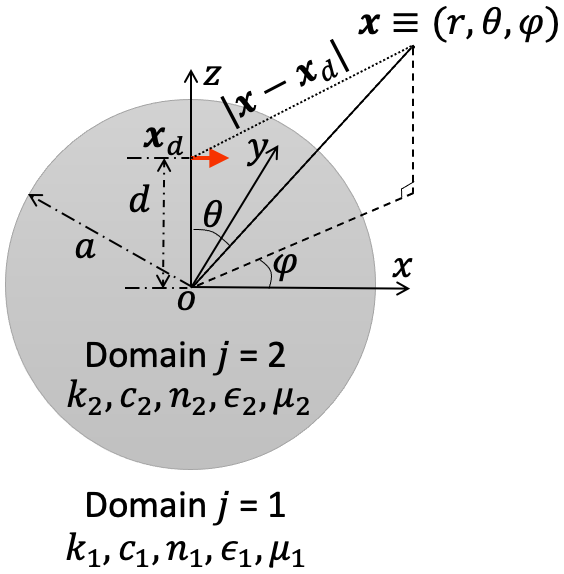}} 
\caption{Sketch of the calculation model for the internal and external electromagnetic fields driven by an electric dipole embedded in a dielectric sphere.}  \label{Fig:app}
\end{figure}

To obtain the electromagnetic field radiated from a single NV centre in a spherical diamond particle to the external domain, it is convenient to use the spherical coordinate system, $(r,\theta,\varphi)$, which origin is at the centre of the diamond particle. As shown in Fig.~\ref{Fig:app}, we assign the symmetric axis is along the $z$-axis which is the polar angle $\theta$ measured from. The equivalent electric dipole for the NV centre is positioned along the axis of symmetry at $d \bs{e}_z =  d \cos{\theta}\bs{e}_r - d \sin{\theta}\bs{e}_{\theta}$. Two situations are considered separately: (i) the vertical dipole when the dipole moment direction is along the symmetric axis ($z$-axis) as $\bs{p} = p \bs{e}_{z} = p \cos{\theta}\bs{e}_r - p \sin{\theta}\bs{e}_{\theta}$ as shown in Fig.~\ref{Fig:app} (a) and detailed in Sec.~\ref{sec:appvertical} and (ii) the horizontal dipole when the dipole moment direction is perpendicular to $z$-axis as $\bs{p} = p \bs{e}_x = p \sin{\theta}\cos{\varphi}\bs{e}_r + p \cos{\theta}\cos{\varphi}\bs{e}_{\theta}-p \sin{\varphi} \bs{e}_{\varphi} $ as shown in Fig.~\ref{Fig:app} (b) and detailed in Sec.~\ref{sec:apphorizontal}. Here, $\bs{e}_r, \, \bs{e}_{\theta}, \, \bs{e}_{\varphi}$ are the unit vector along $r, \, \theta, \, \varphi$ direction in the spherical coordinate system, respectively. All the other dipole location and polarisation scenarios, such as the cases presented in the main text, can be easily obtained through coordinate rotation and linear superposition from the above two basic cases.  

In the spherical coordinate system, the Maxwell's equations in Eq.~(\ref{eq:Maxwelleq}) are in the form of 
\begin{subequations}
    \begin{align}
        \frac{1}{r\sin{\theta}} \left[\frac{\partial}{\partial{\theta}}(E^{j}_{\varphi} \sin{\theta}) - \frac{\partial{E^{j}_{\theta}}}{\partial{\varphi}}\right] = \rmi \omega \mu_0 \mu_j H^{j}_{r}, \label{eq:MaxwellHr}\\
        \frac{1}{r} \left[ \frac{1}{\sin{\theta}} \frac{\partial E^{j}_{r}}{\partial {\varphi}} -\frac{\partial}{\partial r} (r E^{j}_{\varphi})\right] = \rmi \omega \mu_0 \mu_j H^{j}_{\theta}, \label{eq:MaxwellHq} \\
        \frac{1}{r} \left[\frac{\partial}{\partial r} (r E^{j}_{\theta}) - \frac{\partial E^{j}_{r}}{\partial \theta}\right] = \rmi \omega \mu_0 \mu_j H^{j}_{\varphi}; \label{eq:MaxwellHf}
    \end{align}
\end{subequations}
\begin{subequations}
    \begin{align}
        \frac{1}{r\sin{\theta}} \left[\frac{\partial}{\partial{\theta}}(H^{j}_{\varphi} \sin{\theta}) - \frac{\partial{H^{j}_{\theta}}}{\partial{\varphi}}\right] =-\rmi \omega \epsilon_0 \epsilon_j E^{j}_{r}, \label{eq:MaxwellEr}\\
        \frac{1}{r} \left[ \frac{1}{\sin{\theta}} \frac{\partial H^{j}_{r}}{\partial {\varphi}} -\frac{\partial}{\partial r} (r H^{j}_{\varphi})\right] =-\rmi \omega \epsilon_0 \epsilon_j E^{j}_{\theta},\label{eq:MaxwellEq} \\
        \frac{1}{r} \left[\frac{\partial}{\partial r} (r H^{j}_{\theta}) - \frac{\partial H^{j}_{r}}{\partial \theta}\right] =-\rmi \omega \epsilon_0 \epsilon_j E^{j}_{\varphi}.\label{eq:MaxwellEf}
    \end{align}
\end{subequations}
In the above equation, the continuity equations of the electric and magnetic fields are not given as they are satisfied straightforwardly when the Mie solution procedure is used, as demonstrated below. 

Before we solve for the reflection and radiation electromagnetic fields in Domain 1 and 2, we need to write the fields due to the electric dipole in the spherical coordinate system. From Eq.~(\ref{eq:EMfielddipole}), we have
\begin{subequations}\label{eq:EMfielddipoleGreenfun}
    \begin{align}
        \bs{E}^{d} &= \frac{1}{4\pi \epsilon_0 \epsilon_2}\frac{\exp( \rmi k_2 r_d)}{ r_d^3}\Bigg\{ (-k_2^2 r_d^2 - 3 \rmi k_2 r_d +3) \frac{\boldsymbol r_d \cdot \boldsymbol p}{r_d^2}\boldsymbol r_d \nonumber \\
        &\qquad \qquad \qquad \qquad   + (k_2^2 r_d^2 + \rmi k_2 r_d -1) \boldsymbol p \Bigg\} \nonumber \\
        & = \frac{1}{4\pi \epsilon_0 \epsilon_2} \left[ \bs{p} \nabla^2 G(\bs{x},\bs{x}_d)  - (\bs{p}\cdot \nabla) \nabla G(\bs{x},\bs{x}_d) \right], \\
        \bs{H}^{d} &= \frac{\omega k_2}{4\pi}[\bs{r}_d \times \bs{p}] \left( \frac{1}{r_d} - \frac{1}{\rmi k_2 r_d^2} \right) \frac{\exp( \rmi k_2 r_d)}{r_d} \nonumber \\
        & =-\frac{\rmi \omega }{4 \pi} \left[ \nabla G(\bs{x},\bs{x}_d) \times \bs{p} \right]
    \end{align}
\end{subequations}
where $G(\bs{x},\bs{x}_d)$ is the Green's function for the Helmholtz equation as
\begin{align}
G(\bs{x},\bs{x}_d) = \frac{\exp{(\rmi k_2 |\bs{x}-\bs{x}_d|)}}{|\bs{x}-\bs{x}_d|}.
\end{align}
As shown in Fig.~\ref{Fig:app}, $r_d \equiv |\bs{x}-\bs{x}_d| = \sqrt{r^2 + d^2 - 2 r d \cos{\theta}}$ based on the cosine theorem.  In this case, the free space Green's function for the Helmholtz equation can be rewritten in terms of $(r, \, d, \, \theta)$ and asymptotically represented in terms of free spherical multipolar waves, respectively, as 
\begin{align}
    G(\bs{x},\bs{x}_d) \equiv & G(r,\theta;d) \nonumber  \\
     = & \frac{\exp{(\rmi k_2 \sqrt{r^2 + d^2 - 2 r d \cos{\theta}})}}{\sqrt{r^2 + d^2 - 2 r d \cos{\theta}}} \label{eq:GFrdq}  \\
    =& \rmi k_2 \sum^{N}_{n=0} (2n + 1) h_{n}^{(1)} (k_2 r_{>}) j_n(k_2 r_{<}) P_{n}(\cos{\theta}) \label{eq:GFBessel}
\end{align}
where $r_{>} \equiv \max(|\bs{x}|, \, d)$, $r_{<} \equiv \min(|\bs{x}|, \, d)$, and $N = k_1a + 4(k_1a)^{1/3} + 2 $ is the transacted number for the summation~\cite{Bohren1998}. Introducing Eq.~(\ref{eq:GFrdq}) or Eq.~(\ref{eq:GFBessel}) into Eq.~(\ref{eq:EMfielddipoleGreenfun}) and using the vector calculus formulae in the spherical coordinate system, the fields due to the electric dipole in the spherical coordinate system are obtained.

\subsection{Vertical electric dipole}\label{sec:appvertical}
Let us firstly consider to solve for the electromagnetic fields as the case illustrated in Fig.~\ref{Fig:app} (a). Introducing Eq.~(\ref{eq:GFrdq}) into Eq.~(\ref{eq:EMfielddipoleGreenfun}) and using the vector calculus formulae in the spherical coordinate system, the electric and magnetic fields induced by a vertical electric dipole, when $\bs{p} = p \cos{\theta}\bs{e}_r - p \sin{\theta}\bs{e}_{\theta}$, are
\begin{subequations}\label{eq:EMfieldVdip}
    \begin{align}
        E^{d}_{r} & = \frac{1}{4\pi \epsilon_0 \epsilon_2}\frac{p}{d} \left\{ \frac{\partial^2 \left[ r\, G(r,\theta;d)\right]}{\partial r^2 } + k_2^2\, r\, G(r,\theta;d) \right\}, \\
        E^{d}_{\theta} & = \frac{1}{4\pi \epsilon_0 \epsilon_2}\frac{p}{r d} \frac{\partial^2 \left[ r\, G(r,\theta;d)\right]}{\partial r \partial \theta}, \\
        E^{d}_{\varphi} & = 0; \\
        H^{d}_{r} & = 0, \\
        H^{d}_{\theta} & = 0, \\
        H^{d}_{\varphi} & =-\frac{\rmi \omega }{4 \pi} \frac{p}{d}\frac{\partial G(r,\theta;d)}{\partial \theta}.
    \end{align}
\end{subequations}

Based on the Mie theory~\cite{Mie1908} by using Debye potentials $u$ and $v$ that satisfy the Helmholtz equation~\cite{vandeHulst1957, Bohren1998}, we can write the electric and magnetic fields as
\begin{subequations}\label{eq:em_EH_MN}
    \begin{align}
        \bs{E} &= E_0\left(\bs{M}_{v} - \rmi \bs{N}_{u}\right), \\
        \bs{H} &= E_0\sqrt{\frac{\epsilon_0\epsilon_r}{\mu_0 \mu_r}} (-\rmi \bs{N}_v - \bs{M}_u)
    \end{align}
\end{subequations}
where $E_0 = p/(4\pi\epsilon_0 a^3)$, 
\begin{align}
        & \bs{M}_{u} = \nabla \times (\bs{r} u), \nonumber \\
        & \bs{M}_{v} = \nabla \times (\bs{r} v), \nonumber \\
        & \nabla \times \bs{M}_{u} = \omega (\epsilon_0\epsilon_r\mu_0 \mu_r)^{\frac{1}{2}} \bs{N}_u, \nonumber \\
        & \nabla \times \bs{M}_{v} = \omega (\epsilon_0\epsilon_r\mu_0 \mu_r)^{\frac{1}{2}} \bs{N}_v, \nonumber \\
        & \nabla \times \bs{N}_{u} = \omega (\epsilon_0\epsilon_r\mu_0 \mu_r)^{\frac{1}{2}} \bs{M}_u, \nonumber \\
        & \nabla \times \bs{N}_{v} = \omega (\epsilon_0\epsilon_r\mu_0 \mu_r)^{\frac{1}{2}} \bs{M}_v, \nonumber \\
        & \nabla \times \bs{E} = \rmi \omega \mu_0 \mu_r \bs{H}, \nonumber \\
        & \nabla \times \bs{H}  =-\rmi \omega \epsilon_0\epsilon_r \bs{E}.
\end{align}
The full components of $\bs{M}_{u}$ and $\bs{N}_{u}$ are, respectively,
\begin{subequations}\label{eq:MNu}
    \begin{align}
        M_{u_{r}} &= 0, \qquad M_{u_{\theta}} = \frac{1}{r\sin{\theta}} \frac{\partial (ru)}{\partial \varphi}, \nonumber \\
        M_{u_{\varphi}} & =-\frac{1}{r} \frac{\partial(ru)}{\partial \theta}; \\
        N_{u_{r}} & = \frac{1}{k} \frac{\partial^2(ru)}{\partial r^2} + k  r u, \qquad N_{u_{\theta}}  = \frac{1}{k r} \frac{\partial^2(ru)}{\partial r \partial \theta}, \nonumber \\
        N_{u_{\varphi}} & = \frac{1}{k r \sin\theta} \frac{\partial^2(ru)}{\partial r \partial \varphi}.
    \end{align}
\end{subequations}
The above formulations can also be used to get the components of $\bs{M}_{v}$ and $\bs{N}_{v}$ when potential $u$ is replaced by potential $v$. 

As Debye potentials $u$ and $v$ satisfy the Helmholtz equation, let us consider a scalar wave equation for function $\phi$ with wavenumber $k$:
\begin{align}\label{eq:waveeq}
    \nabla^2 \phi + k^2 \phi = 0
\end{align}
where $\phi$ represents either potential $u$ or $v$. Eq.~(\ref{eq:waveeq}) is variable separable in the spherical coordinate system, and its elementary solutions are
\begin{subequations}\label{eq:waveeq_sol}
    \begin{align}
    \phi_{(l,n)} = &\sum_{n=0}^{\infty} \sum_{l=-n}^{l=n} C_{l,n} \cos{(l \varphi)} P_{n}^{l}(\cos{\theta}) z_{n}(kr), \\
    \phi_{(l,n)} = &\sum_{n=0}^{\infty} \sum_{l=-n}^{l=n} D_{l,n} \sin{(l \varphi)} P_{n}^{l}(\cos{\theta}) z_{n}(kr)
    \end{align}
\end{subequations}
where $l$ and $n$ are integers ($n \ge l \ge 0$), $P_{n}^{l}(\cos{\theta})$ is an associated Legendre polynomial, and $z_{n}(kr)$ is the spherical Bessel function of any kind. The following rules are applied to determine the choice of function $z_{n}(kr)$. In the bounded domain with origin within it, $j_{n}(kr)$, the spherical Bessel function of the first kind, is used as $j_{n}(kr)$ is finite at origin. In the bounded domain excluding origin, both $j_{n}(kr)$ and $y_n(kr)$, the spherical Bessel functions of the first and second kinds, are needed. In the unbounded external domain, for the scattered or radiation field, $h_{n}^{(1)} = j_{n}(kr) + \rmi y_n(kr)$ is used as $\rmi k h_{n}^{(1)} \sim \rmi^{n} \exp{(\rmi k r)}/r$. %A good reference for all the special functions mentioned above is https://dlmf.nist.gov/. So useful recurrence relations and derivatives are listed below.

It is worth noting that the two Debye potentials, $u$ and $v$, correspond to $\cos(l\varphi)$ and $\sin(l\varphi)$ formulations in Eq.~(\ref{eq:waveeq_sol}), respectively. Nevertheless, according to Eq.~(\ref{eq:EMfieldVdip}), the fields driven by a vertical electric dipole in a sphere do not depend on $\varphi$. As such, only terms with $l=0$ in Eq.~(\ref{eq:waveeq_sol}) are needed, which means only one potential is needed for each domain. Let us use potential $u$:
\begin{subequations}\label{eq:em_dip_u}
    \begin{align}
        u_{(0,n)}^{1} &= \sum_{n=0}^{N} C^{1}_{(0,n)}  P_{n}(\cos{\theta}) h_{n}^{(1)}(k_1 r),\\
        u_{(0,n)}^{2} &= \sum_{n=0}^{N} C^{2}_{(0,n)}  P_{n}(\cos{\theta}) j_{n}(k_2 r)
    \end{align}
\end{subequations}
for the external and internal domain, respectively, where the $C^{1}_{(0,n)}$ and $C^{2}_{(0,n)}$ are determined by the boundary conditions. Introducing Eq.~(\ref{eq:em_dip_u}) into Eq.~(\ref{eq:em_EH_MN}) and using Eq.~(\ref{eq:MNu}), we obtain
\begin{subequations}\label{eq:EHdm1fieldVdip}
\begin{align}
    E_{r}^{1} = & E_0 \,  \sum_{n=0}^{N}C^{1}_{(0,n)} (-\rmi) \frac{n (n+1)}{k_1 r}h^{(1)}_{n}(k_1 r)  P_{n}(\cos{\theta}), \\
    E_{\theta}^{1} = & E_{0} \, \sum_{n=0}^{N} C^{1}_{(0,n)} (-\rmi) \frac{1}{k_1 r}\Big[ (n+1) h^{(1)}_{n}(k_1r) \nonumber \\
    & \qquad \qquad \qquad - k_1 r h^{(1)}_{n+1}(k_1 r) \Big] P_{n}^{1}(\cos\theta), \\
    H_{\varphi}^{1} = &  \frac{E_0}{\omega \mu_0} \sum_{n=0}^{N} C^{1}_{(0,n)} \frac{k_1}{\mu_1} h^{(1)}_{n}(k_1 r)  P^{1}_{n}(\cos{\theta}), 
\end{align}
\end{subequations}
and $E_{\varphi}^{1}  =  H_{r}^{1} = H_{\theta}^{1} = 0$.
Also,
\begin{subequations}\label{eq:EHdm2fieldVdip}
\begin{align}
    E_{r}^{2} = & E_0 \,  \sum_{n=0}^{N}C^{2}_{(0,n)} (-\rmi) \frac{n (n+1)}{k_2 r}j_{n}(k_2 r)  P_{n}(\cos{\theta}), \\
    E_{\theta}^{2} = & E_{0} \, \sum_{n=0}^{N} C^{2}_{(0,n)} (-\rmi) \frac{1}{k_2 r}\Big[ (n+1) j_{n}(k_2r) \nonumber \\
    & \qquad \qquad \qquad- k_2 r j_{n+1}(k_2 r) \Big] P_{n}^{1}(\cos\theta), \\
    H_{\varphi}^{2} = & \frac{E_0}{\omega \mu_0} \sum_{n=0}^{N} C^{2}_{(0,n)} \frac{k_2}{\mu_2} j_{n}(k_2 r)  P^{1}_{n}(\cos{\theta}), 
\end{align}
\end{subequations}
and $E_{\varphi}^{2} = H_{r}^{2} = H_{\theta}^{2} = 0$.

To get $C^{1}_{(0,n)}$ and $C^{2}_{(0,n)}$, the boundary conditions across the sphere surface:
\begin{subequations}\label{eq:EHBCVdip}
    \begin{align}
        E_{\theta}^{1} = E_{\theta}^{2} + E_{\theta}^{d} \qquad \text{when} \quad r = a, \\
        H_{\varphi}^{1} = H_{\varphi}^{2} + H_{\varphi}^{d}  \qquad \text{when} \quad r = a
    \end{align}
\end{subequations} 
are used. Introducing Eq.~(\ref{eq:GFBessel}) into Eq.~(\ref{eq:EMfieldVdip}) and setting $r=a$, we have
\begin{subequations}\label{eq:EMfieldVdipBC}
    \begin{align}
        E^{d}_{\theta}\Big|_{r=a}  = \frac{1}{4\pi \epsilon_0 \epsilon_2}\frac{p}{a d} \rmi k_2 & \sum^{N}_{n=0}  (2n + 1)  j_n(k_2 d) P^{1}_{n}(\cos{\theta}) \nonumber \\
        & \quad \times \frac{\rmd{\left[ r h_{n}^{(1)} (k_2 r)\right] }}{\rmd r}\Bigg|_{r=a}, \\
        H^{d}_{\varphi}\Big|_{r=a}  =-\frac{\rmi \omega }{4 \pi} \frac{p}{d}\rmi k_2 & \sum^{N}_{n=0} (2n + 1) h_{n}^{(1)} (k_2 a) j_n(k_2 d) \nonumber \\
        &\quad \times P^{1}_{n}(\cos{\theta}).
    \end{align}
\end{subequations}
Letting $r=a$ in Eqs.~(\ref{eq:EHdm1fieldVdip}) and (\ref{eq:EHdm2fieldVdip}), and introducing the results and Eq.~(\ref{eq:EMfieldVdipBC}) into Eq.~(\ref{eq:EHBCVdip}), we obtain a $2\times 2$ linear system to solve for the unknown coefficients $C^{1}_{(0,n)}$ and $C^{2}_{(0,n)}$ which can be then introduced into Eqs.~(\ref{eq:EHdm1fieldVdip}) and (\ref{eq:EHdm2fieldVdip}) to calculate the fields inside, outside the sphere and on the sphere surface. 

\begin{figure*}[t]
\centering{}
\subfloat[Vertical dipole]{ \includegraphics[width=0.32\textwidth]{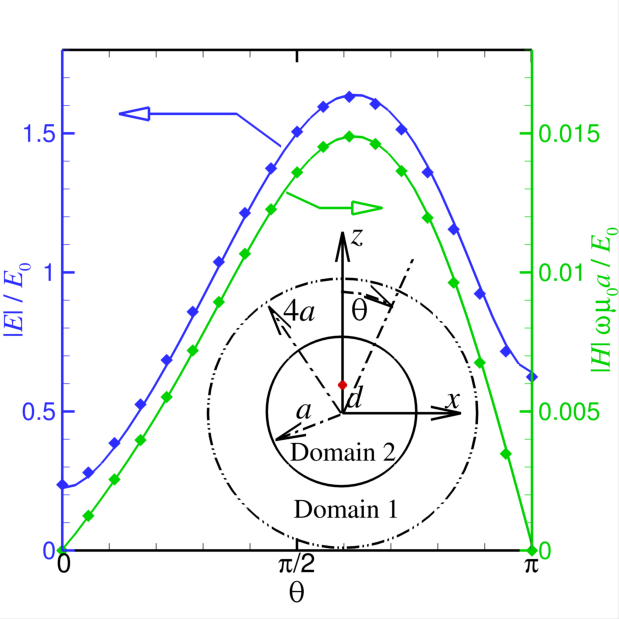}} \qquad \qquad \qquad
\subfloat[Horizontal dipole]{ \includegraphics[width=0.32\textwidth]{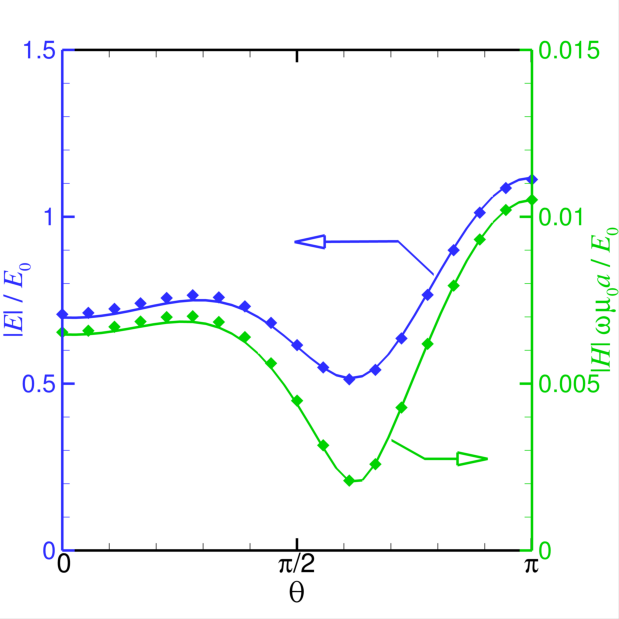}}
\caption{Good agreement has been found for the electromagnetic fields between the results obtained by the asymptotic approximations shown in Section~\ref{sec:appsolution} (solid lines), by the in-house built field only surface integral method~\cite{Sun2022} (symbols) when $a=200$ nm, $d=50$ nm, $n_1 = 1$, $n_2=2.4$, $E_0 = p/(4\pi \epsilon_0 a^3)$ and $\lambda = 708$ nm: (a) a vertical electric dipole and (b) a horizontal electric dipole. The curves plotted are the magnitudes of the electromagnetic fields along the circle concentric with the spherical particle with radius as $4a$ on the $xz$ plane (as shown in the inset of Fig.~\ref{Fig:AppComp}a).  }  \label{Fig:AppComp}
\end{figure*}

\subsection{Horizontal electric dipole}\label{sec:apphorizontal}
Let us turn to solve for the electromagnetic fields as the case illustrated in Fig.~\ref{Fig:app} (b). Introducing Eq.~(\ref{eq:GFrdq}) into Eq.~(\ref{eq:EMfielddipoleGreenfun}) and using the vector calculus formulae in the spherical coordinate system, the radial components of the electric and magnetic fields induced by a horizontal electric dipole when $\bs{p} = p \sin{\theta}\cos{\varphi}\bs{e}_r + p \cos{\theta}\cos{\varphi}\bs{e}_{\theta}-p \sin{\varphi} \bs{e}_{\varphi} $ are
% \begin{subequations}\label{eq:EMfieldHdip}
%     \begin{align}
%         E^{d}_{r} & = \frac{1}{4\pi \epsilon_0 \epsilon_2}\frac{p}{d} \left\{ \frac{\partial^2 \left[ r\, G(r,\theta;d)\right]}{\partial r^2 } + k_2^2\, r\, G(r,\theta;d) \right\}, \\
%         E^{d}_{\theta} & = \frac{1}{4\pi \epsilon_0 \epsilon_2}\frac{p}{r d} \frac{\partial^2 \left[ r\, G(r,\theta;d)\right]}{\partial r \partial \theta}, \\
%         E^{d}_{\varphi} & = 0; \\
%         H^{d}_{r} & =  \frac{\rmi \omega }{4 \pi} p \frac{1}{r}\frac{\partial G(r,\theta;d)}{\partial \theta} \sin{\varphi}, \\
%         H^{d}_{\theta} & =-\frac{\rmi \omega }{4 \pi} p \frac{\partial G(r,\theta;d)}{\partial r} \sin{\varphi}, \\
%         H^{d}_{\varphi} & = \frac{\rmi \omega }{4 \pi} p \frac{\partial G(r,\theta;d)}{\partial d} \cos{\varphi}..
%     \end{align}
% \end{subequations}
\begin{subequations}\label{eq:EMrfieldHdip}
    \begin{align}
        E^{d}_{r} & =-\frac{1}{4\pi \epsilon_0 \epsilon_2}\frac{p}{r} \frac{\partial}{\partial \theta}\left\{ \frac{\partial \left[ G(r,\theta;d)\right]}{\partial d } + \frac{1}{d} G(r,\theta;d) \right\} \cos{\varphi}, \\
        H^{d}_{r} & =  \frac{\rmi \omega }{4 \pi} \frac{p}{r}\frac{\partial G(r,\theta;d)}{\partial \theta} \sin{\varphi}.
    \end{align}
\end{subequations}

Following the same solution procedure shown in the previous section and considering that the electromagnetic fields given in Eq.~(\ref{eq:EMrfieldHdip}) are functions of $\sin{\varphi}$ and $\cos(\varphi)$, only the terms when $l=1$ from the elementary solutions in Eq.~(\ref{eq:waveeq_sol}) are needed for the Debye potentials. As such, the following Debye potentials 
\begin{subequations}\label{eq:em_hdip_uv}
    \begin{align}
        u_{(1,n)}^{1} = &\cos{\varphi} \sum_{n=1}^{N} C^1_{(1,n)}  h^{(1)}_{n}(k_1 r) P_{n}^{1}(\cos{\theta}),  \\
        v_{(1,n)}^{1} =-&\sin{\varphi} \sum_{n=1}^{N} D^1_{(1,n)}  h^{(1)}_{n}(k_1 r) P_{n}^{1}(\cos{\theta}); \\
        u_{(1,n)}^{2} = &\cos{\varphi} \sum_{n=1}^{N} C^2_{(1,n)}  j_{n}(k_1 r) P_{n}^{1}(\cos{\theta}), \\
        v_{(1,n)}^{2} =-&\sin{\varphi} \sum_{n=1}^{N} D^2_{(1,n)}  j_{n}(k_1 r) P_{n}^{1}(\cos{\theta})
    \end{align}
\end{subequations}
for the external (with superscript 1) and internal domain (with superscript 2), respectively, are used where $C^1_{(1,n)}, \, D^1_{(1,n)}, \, C^2_{(1,n)}, \, D^2_{(1,n)}$ are unknowns to be determined via boundary conditions.

Introducing Eq.~(\ref{eq:em_hdip_uv}) into Eq.~(\ref{eq:em_EH_MN}) and using Eq.~(\ref{eq:MNu}), we obtain
\begin{widetext}
\begin{subequations}\label{eq:E1rqf_Hdip}
    \begin{align}
        E_{r}^{1} = &\quad E_0 \,  \cos{\varphi} \, \sum_{n=1}^{N}C^1_{(1,n)} (-\rmi) \frac{n(n+1)}{k_1 r}h^{(1)}_{n}(k_1 r)  P^{1}_{n}(\cos{\theta}),  \\
        E_{\theta}^{1} = &\quad  E _{0} \cos{\varphi} \, \sum_{n=1}^{N} C^1_{(1,n)}(-\rmi) \frac{1}{k_1 r}\left[(n+1) h^{(1)}_{n}(k_1 r) - k_1 r h^{(1)}_{n+1}(k_1 r) \right] \frac{ \rmd P_{n}^{1}(\cos\theta)}{\rmd \theta} \nonumber \\
        & + E_{0} \cos{\varphi} \, \sum_{n=1}^{N} D^1_{(1,n)} (-1) h^{(1)}_{n}(k_1 r) \frac{P^{1}_{n}(\cos{\theta})} {\sin{\theta}}, \\
        E_{\varphi}^{1} = & \quad E_{0} \sin{\varphi}\, \sum_{n=1}^{N} C^1_{(1,n)} (\rmi) \frac{1}{k_1 r}\left[(n+1) h^{(1)}_{n}(k_1 r) - k_1 r h^{(1)}_{n+1}(k_1 r) \right]  \frac{P^{1}_{n}(\cos{\theta})} {\sin{\theta}} \nonumber \\
        & + E_{0} \sin{\varphi} \, \sum_{n=1}^{N} D^1_{(1,n)} h^{(1)}_{n}(k_1 r) \frac{ \rmd P_{n}^{1}(\cos\theta)}{\rmd \theta}; \label{eq:Efdm1Hdip}
    \end{align}
\end{subequations}
\end{widetext}
% and
\begin{widetext}
\begin{subequations}
    \begin{align}
        H_{r}^{1} = &\quad \frac{E_0}{\omega \mu_0}  \sin{\varphi} \, \sum_{n=1}^{N}D^1_{(1,n)} (\rmi) \frac{k_1}{\mu_1} \frac{n(n+1)}{k_1 r}h^{(1)}_{n}(k_1 r)  P^{1}_{n}(\cos{\theta}), \\
        H_{\theta}^{1} = & \quad \frac{E_0}{\omega \mu_0}  \sin{\varphi} \, \sum_{n=1}^{N} C^1_{(1,n)} \frac{k_1}{\mu_1} h^{(1)}_{n}(k_1 r) \frac{P^{1}_{n}(\cos{\theta})} {\sin{\theta}} \nonumber \\
        & + \frac{E_0}{\omega \mu_0} \sin{\varphi} \,  \, \sum_{n=1}^{N} D^1_{(1,n)} (\rmi) \frac{k_1}{\mu_1} \frac{1}{k_1 r}\left[(n+1) h^{(1)}_{n}(k_1 r) - k_1 r h^{(1)}_{n+1}(k_1 r) \right] \frac{ \rmd P_{n}^{1}(\cos\theta)}{\rmd \theta}, \label{eq:Hqdm1Hdip} \\
        H_{\varphi}^{1} = & \quad \frac{E_0}{\omega \mu_0}\cos{\varphi} \, \sum_{n=1}^{N} C^1_{(1,n)} \frac{k_1}{\mu_1} h^{(1)}_{n}(k_1 r) \frac{ \rmd P_{n}^{1}(\cos\theta)}{\rmd \theta} \nonumber \\
        & + \frac{E_0}{\omega \mu_0} \cos{\varphi}\, \sum_{n=1}^{N} D^1_{(1,n)} (\rmi)\frac{k_1}{\mu_1} \frac{1}{k_1 r}\left[ (n+1) h^{(1)}_{n}(k_1 r) - k_1 rh^{(1)}_{n+1}(k_1 r) \right]  \frac{P^{1}_{n}(\cos{\theta})} {\sin{\theta}}.
    \end{align}
\end{subequations}
\end{widetext}
Also
\begin{widetext}
\begin{subequations}
    \begin{align}
        E_{r}^{2} = &E_0 \,  \cos{\varphi} \, \sum_{n=1}^{N}C^2_{(1,n)} (-\rmi) \frac{n(n+1)}{k_2 r}j_{n}(k_2 r)  P^{1}_{n}(\cos{\theta}),  \\
        E_{\theta}^{2} = &\quad  E _{0} \cos{\varphi} \, \sum_{n=1}^{N} C^2_{(1,n)}(-\rmi) \frac{1}{k_2 r}\left[(n+1) j_{n}(k_2 r) - k_2 r j_{n+1}(k_2 r) \right] \frac{ \rmd P_{n}^{1}(\cos\theta)}{\rmd \theta} \nonumber \\
        & + E_{0} \cos{\varphi} \, \sum_{n=1}^{N} D^2_{(1,n)} (-1) j_{n}(k_2 r) \frac{P^{1}_{n}(\cos{\theta})} {\sin{\theta}}, \\
        E_{\varphi}^{2} = & \quad E_{0} \sin{\varphi}\, \sum_{n=1}^{N} C^2_{(1,n)} (\rmi) \frac{1}{k_2 r}\left[(n+1) j_{n}(k_2 r) - k_2 r j_{n+1}(k_2 r) \right]  \frac{P^{1}_{n}(\cos{\theta})} {\sin{\theta}} \nonumber \\
        & + E_{0} \sin{\varphi} \, \sum_{n=1}^{N} D^2_{(1,n)} j_{n}(k_2 r) \frac{ \rmd P_{n}^{1}(\cos\theta)}{\rmd \theta}; \label{eq:Efdm2Hdip}
    \end{align}
\end{subequations}
\end{widetext}
% and
\begin{widetext}
\begin{subequations}\label{eq:H2rqf_Hdip}
    \begin{align}
        H_{r}^{1} = & \frac{E_0}{\omega \mu_0}  \sin{\varphi} \, \sum_{n=1}^{N}D^2_{(1,n)} (\rmi) \frac{k_2}{\mu_2} \frac{n(n+1)}{k_2 r}j_{n}(k_2 r)  P^{1}_{n}(\cos{\theta}), \\
        H_{\theta}^{1} = & \quad \frac{E_0}{\omega \mu_0}  \sin{\varphi} \, \sum_{n=1}^{N} C^2_{(1,n)} \frac{k_2}{\mu_2} j_{n}(k_2 r) \frac{P^{1}_{n}(\cos{\theta})} {\sin{\theta}} \nonumber \\
        & + \frac{E_0}{\omega \mu_0} \sin{\varphi} \,  \, \sum_{n=1}^{N} D^2_{(1,n)} (\rmi) \frac{k_2}{\mu_2} \frac{1}{k_2 r}\left[(n+1) j_{n}(k_2 r) - k_2 r j_{n+1}(k_2 r) \right] \frac{ \rmd P_{n}^{1}(\cos\theta)}{\rmd \theta},  \label{eq:Hqdm2Hdip} \\
        H_{\varphi}^{1} = & \quad \frac{E_0}{\omega \mu_0}\cos{\varphi} \, \sum_{n=1}^{N} C^2_{(1,n)} \frac{k_2}{\mu_2} j_{n}(k_2 r) \frac{ \rmd P_{n}^{1}(\cos\theta)}{\rmd \theta} \nonumber \\
        & + \frac{E_0}{\omega \mu_0} \cos{\varphi}\, \sum_{n=1}^{N} D^2_{(1,n)} (\rmi)\frac{k_2}{\mu_2} \frac{1}{k_2 r}\left[ (n+1) j_{n}(k_2 r) - k_2 rj_{n+1}(k_2 r) \right]  \frac{P^{1}_{n}(\cos{\theta})} {\sin{\theta}}.
    \end{align}
\end{subequations}
\end{widetext}

Once the coefficients $C^1_{(1,n)}, \, D^1_{(1,n)}, \, C^2_{(1,n)}, \, D^2_{(1,n)}$ are found, the electromagnetic fields in both domains are determined. To get those coefficients, the boundary conditions for the tangential components of the electric and magnetic fields due to electric dipole on the sphere surface when $r=a$ are need, which can be found by using the Maxwell's equations and the radial components in Eq.~(\ref{eq:EMrfieldHdip})~\cite{Margetis2002}. Introducing Eq.~(\ref{eq:MaxwellHq}) in Eq.~(\ref{eq:MaxwellEf}), for the sphere domain, we get
\begin{align}\label{eq:EfHdipwaveEq}
    \frac{\partial^2}{\partial r^2} (r E^{d}_{\varphi}) + k_2^2 (r E^{d}_{\varphi}) = \frac{1}{\sin{\theta}} \frac{\partial}{\partial r} \left( \frac{\partial E^{d}_{r}}{\partial \varphi}\right) - \rmi \omega \mu_0 \mu_2 \frac{\partial H^{d}_{r}}{\partial \theta}.
\end{align}
The right-hand-side of Eq.~(\ref{eq:EfHdipwaveEq}) can be obtained by using the results from introducing Eq.~(\ref{eq:GFBessel}) into Eq.~(\ref{eq:EMrfieldHdip}):
\begin{align}\label{eq:ErrhdHdip}
    &\frac{1}{\sin{\theta}} \frac{\partial}{\partial r} \left( \frac{\partial E^{d}_{r}}{\partial \varphi}\right) \nonumber \\
    =& \frac{p\sin{\varphi}}{4\pi \epsilon_0 \epsilon_2} \frac{\rmi k_2}{d} \sum^{N}_{n=1} (2n + 1)\frac{\rmd}{\rmd r} \left[\frac{h_{n}^{(1)} (k_2 r)}{r}\right] \nonumber \\
    & \qquad \; \times \left[(n+1)j_n(k_2 d) - k_2 d \, j_{n+1}(k_2 d)\right] \frac{P^{1}_{n}(\cos{\theta})}{\sin{\theta}},
\end{align}
and
\begin{align}\label{eq:HrrhdHdip}
    - \rmi \omega \mu_0 \mu_2 \frac{\partial H^{d}_{r}}{\partial \theta} =  \frac{\rmi p k_2^3\sin{\varphi}}{4\pi \epsilon_0 \epsilon_2}  \sum^{N}_{n=1} & (2n + 1) \frac{h_{n}^{(1)} (k_2 r)}{r}  j_n(k_2 d) \nonumber \\
    &\times \frac{\rmd P^{1}_{n}(\cos{\theta})}{\rmd \theta},
\end{align}
in which the situation for the fields on the sphere surface when $r \rightarrow a > d$ is implied. When comparing the left-hand-side of Eq.~(\ref{eq:EfHdipwaveEq}) and Eqs.~(\ref{eq:ErrhdHdip}) and (\ref{eq:HrrhdHdip}), we notice that we can get the tangential component, $E^{d}_{\varphi}$, by solving the following two ordinary differential equations:
\begin{align}
    \frac{\rmd^2 g_1(r)}{\partial r^2}  + k_2^2 g_1(r) &= \frac{\rmd}{\rmd r} \left[\frac{h_{n}^{(1)} (k_2 r)}{r}\right], \\
    \frac{\rmd^2 g_2(r)}{\partial r^2}  + k_2^2 g_2(r) &=  \frac{h_{n}^{(1)} (k_2 r)}{r}.
\end{align}
The solutions to the above two equations are, respectively, 
\begin{align}
    g_1(r) & = \frac{1}{n(n+1)}\frac{\rmd}{\rmd r} \left[r h_{n}^{(1)} (k_2 r)\right] \nonumber \\
    &= \frac{1}{n(n+1)} \left[(n+1)h_{n}^{(1)}(k_2 r) - k_2 r \, h_{n+1}^{(1)}(k_2 r)\right], \\
    g_2(r) &=  \frac{1}{n(n+1)} r  h_{n}^{(1)}(k_2 r).
\end{align}
As such, 
\begin{widetext}
\begin{align}\label{eq:Efhdip}
  E^{d}_{\varphi} =&\quad \frac{p\sin{\varphi}}{4\pi \epsilon_0 \epsilon_2} \frac{\rmi k_2}{rd} \sum^{N}_{n=1} \frac{2n + 1}{n(n+1)}\frac{\rmd}{\rmd r} \left[r h_{n}^{(1)} (k_2 r)\right] \left[(n+1)j_n(k_2 d) - k_2 d \, j_{n+1}(k_2 d)\right] \frac{P^{1}_{n}(\cos{\theta})}{\sin{\theta}} \nonumber \\
    &+\frac{ p \sin{\varphi}}{4\pi \epsilon_0 \epsilon_2} \rmi k_2^3 \sum^{N}_{n=1} \frac{2n + 1}{n(n+1)} h_{n}^{(1)}(k_2 r)  j_n(k_2 d) \frac{\rmd P^{1}_{n}(\cos{\theta})}{\rmd \theta}.
\end{align}
\end{widetext}

Introducing Eq.~(\ref{eq:GFBessel}) into Eq.~(\ref{eq:EMrfieldHdip}) and substituting that result and Eq.~(\ref{eq:Efhdip}) into Eq.~(\ref{eq:MaxwellHq}), we have
\begin{widetext}
\begin{align}\label{eq:Hqhdip}
    H^{d}_{\theta} = & \quad \frac{1}{\rmi \omega \mu_0 \mu_2} \left[ \frac{1}{r\sin{\theta}} \frac{\partial E^{d}_{r}}{\partial \varphi} - \frac{1}{r}\frac{\partial (r E^{d}_{\varphi})}{\partial r}  \right] \nonumber \\
    = & \quad \frac{p \omega \sin{\varphi} }{4\pi} \frac{1}{rd} \sum^{N}_{n=1} \frac{2n + 1}{n(n+1)} k_2 r h_{n}^{(1)} (k_2 r) \left[(n+1)j_n(k_2 d) - k_2 d \, j_{n+1}(k_2 d)\right]  \frac{P^{1}_{n}(\cos{\theta})}{\sin{\theta}} \nonumber \\
     & - \frac{p \omega \sin{\varphi} }{4\pi} \frac{k_2}{r} \sum^{N}_{n=1} \frac{2n + 1}{n(n+1)} \frac{\rmd}{\rmd r} \left[r h_{n}^{(1)} (k_2 r)\right] j_n(k_2 d) \frac{\rmd P^{1}_{n}(\cos{\theta})}{\rmd \theta}.
\end{align}
\end{widetext}

As the tangential components of the electric and magnetic fields are continuous across the sphere surface, we have
\begin{subequations}\label{eq:EHBCHdip}
    \begin{align}
        E_{\varphi}^{1} = E_{\varphi}^{2} + E_{\varphi}^{d} \qquad \text{when} \quad r = a, \\
        H_{\theta}^{1} = H_{\theta}^{2} + H_{\theta}^{d}  \qquad \text{when} \quad r = a.
    \end{align}
\end{subequations} 
Comparing the expressions in Eqs.~(\ref{eq:Efdm1Hdip}), (\ref{eq:Hqdm1Hdip}), (\ref{eq:Efdm2Hdip}), (\ref{eq:Hqdm2Hdip}) and those in Eqs.~(\ref{eq:Efhdip}), (\ref{eq:Hqhdip}), we obtain a $4\times 4$ linear system to solve for the unknown coefficients $C^1_{(1,n)}$, $D^1_{(1,n)}$, $C^2_{(1,n)}$ and $D^2_{(1,n)}$ that can be introduced back into Eqs.~(\ref{eq:E1rqf_Hdip}) and (\ref{eq:H2rqf_Hdip}) to calculate the fields inside, outside the sphere and on the sphere surface.

In Fig.~\ref{Fig:AppComp}, we showed the electromagnetic fields obtained by the asymptotic approximations detailed in Section~\ref{sec:appsolution}, and compared them with the results gotten by the in-house built field only surface integral method~\cite{Klaseboer2017-NFO, Sun2017-RMF, Sun2020-PEC, Sun2020-DIEL, Sun2022, Klaseboer2022}. Good agreement has been found between the results obtained by the different methods mentioned above.

\bibliography{sample}% Produces the bibliography via BibTeX.

\end{document}